\documentclass[aps,prd,showpacs,onecolumn,preprintnumbers,notitlepage,nofootinbib,tightenlines,10pt]{revtex4-1}
\usepackage{amsmath,bm}
\usepackage{times}
\usepackage{dcolumn}
\usepackage{amsfonts}
\usepackage{amssymb}
\usepackage{braket}
\usepackage{color,graphicx}
\usepackage{epstopdf}
\usepackage{latexsym}
\usepackage[dvipsnames]{xcolor}
\usepackage{slashed}
\usepackage{CJK,upgreek,fancyhdr}
\usepackage{float}
\usepackage{multirow}
\usepackage{hyperref}
\hypersetup{colorlinks,citecolor=nicegreen,linkcolor=nicered,urlcolor=RoyalBlue}
\usepackage{enumitem}
\usepackage{natbib}
\usepackage{relsize}
\usepackage[left=2.5cm,right=2.5cm,top=2.0cm,bottom=2.5cm]{geometry}
\newcommand{\beq}{\begin{eqnarray}}
\newcommand{\eeq}{\end{eqnarray}}
\newcommand{\non}{\nonumber\\ }

\newcommand{\psl}{ P \hspace{-2.4truemm}/ }
\newcommand{\nsl}{ n \hspace{-2.2truemm}/ }
\newcommand{\vsl}{ v \hspace{-2.2truemm}/ }

\def\lsim{ {\ \lower-1.2pt\vbox{\hbox{\rlap{$<$}\lower6pt\vbox{\hbox{$\sim$}
		}}}\ } }
\def\gsim{ {\ \lower-1.2\vbox{\hbox{\rlap{$>$}\lower6pt\vbox{\hbox{$\sim$}
		}}}\ } }

\definecolor{Red}{rgb}{1.,0.,0.}
\definecolor{Blue}{rgb}{0.,0.,1.}
\definecolor{RoyalBlue}{rgb}{0.0,0.14,0.4}
\definecolor{nicered}{rgb}{0.7,0.1,0.2}
\definecolor{nicegreen}{rgb}{0.1,0.4,0.2}

\newcommand{\RyB}[1]{{\color{RoyalBlue}{#1}}}
\def\orcid#1{\kern .08em\href{https://orcid.org/#1}{\includegraphics[keepaspectratio,width=0.76em]{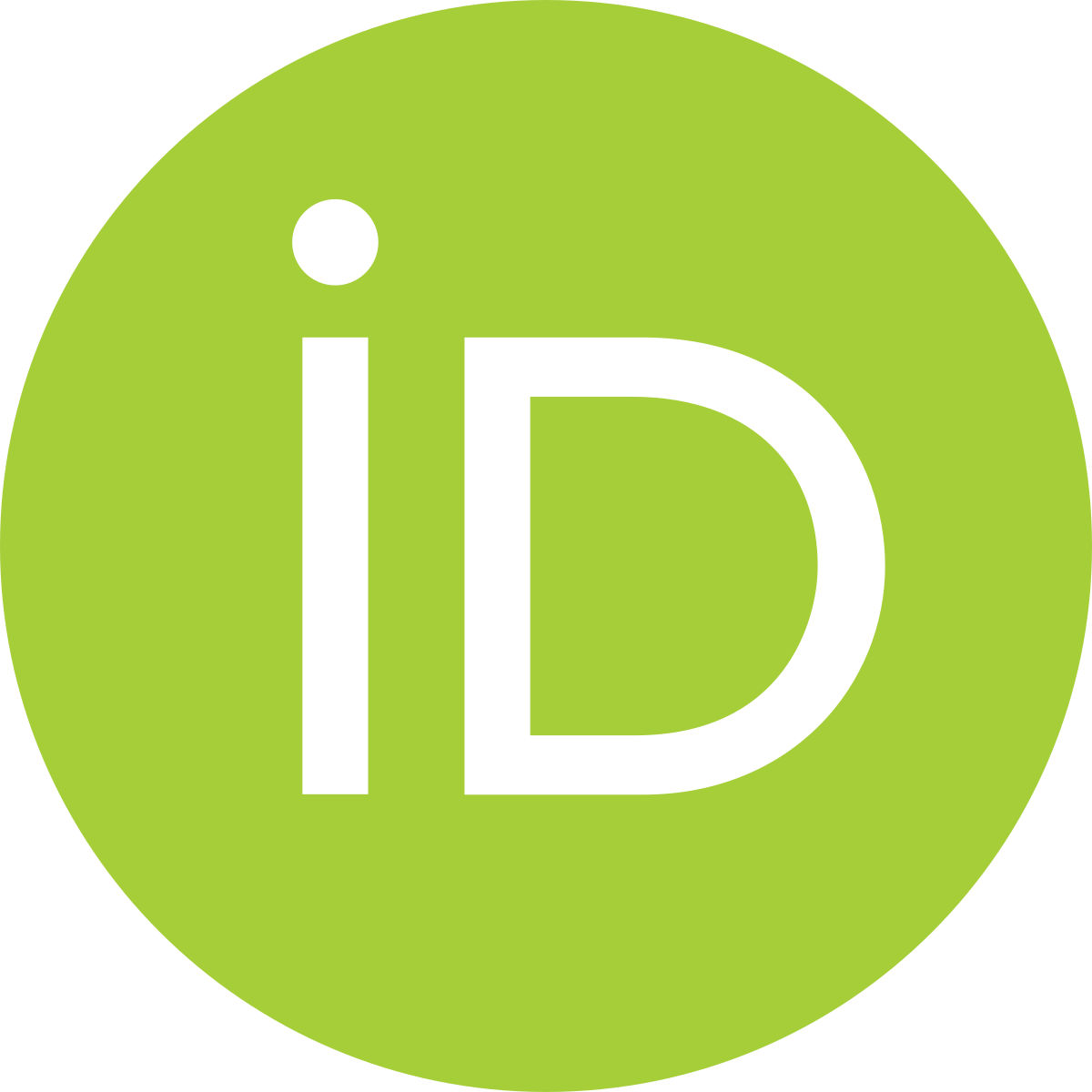}}}

\begin{document}
\title{\boldmath  Branching ratios and {\it CP} asymmetries of $B^0 \to \eta_c f_0$
in the improved perturbative QCD formalism}
\author{Min-Qi~Li\orcid{0009-0006-9738-3424}}
\author{Xin~Liu\orcid{0000-0001-9419-7462}}
\affiliation{Department of Physics,
	Jiangsu Normal University, Xuzhou 221116, China}

\author{Zhi-Tian~Zou\orcid{0000-0002-6985-8174}}
\author{Ying~Li\orcid{0000-0002-1337-7662}}
\affiliation{ Department of Physics, Yantai University, Yantai 264005, China}

\author{Zhen-Jun~Xiao\orcid{0000-0002-4879-209X}}
\affiliation{Department of Physics and Institute of Theoretical Physics,\\
	Nanjing Normal University, Nanjing 210023, China}

\date{\today{}}
	
\begin{abstract}
Motivated by the idea of fragmented scalar glueball, we investigate the decays
$B^0 \to \eta_c f_0$ within the improved perturbative QCD (iPQCD) framework
by including the known next-to-leading order corrections. Here, $B^0$ and
$f_0$ denote the neutral $B_{d,s}^0$ mesons and the light scalar mesons $f_0(500,
980, 1370, 1500)$ under the $q\bar q$ assignment. The {\it CP}-averaged branching
ratios (BRs) and the {\it CP} asymmetries of $B^0 \to \eta_c f_0$ are
evaluated with the $f_0(500)[f_0(1370)]-f_0(980)[f_0(1500)]$ mixing in quark-flavor basis.
For effective comparisons with the near-future measurements, we further
derive the $B^0 \to \eta_c f_0 (\to \pi^+ \pi^-/K^+ K^-)$ BRs under the narrow-width
approximation. ${\rm BR}(B_s^0 \to \eta_c f_0(980) (\to \pi^+ \pi^-))=(2.87^{+1.38}_{-1.29})
\times 10^{-4}$ and ${\rm BR}(B_d^0 \to \eta_c f_0(500)(\to \pi^+\pi^-))/{\rm BR}(B_s^0 \to \eta_c
f_0(980)(\to \pi^+\pi^-)) = (12^{+8}_{-7})\%$ obtained in the iPQCD formalism agree
with the available measurements and/or predictions within uncertainties.
Large BRs of $B_s^0 \to \eta_c f_0(1500)
(\to \pi^+ \pi^-/K^+ K^-)$ and large direct {\it CP} asymmetries of $B^0 \to \eta_c
f_0(1370, 1500)$ are accessible in the LHCb and Belle-II experiments.
The experimental tests of these iPQCD predictions would help us to understand the nature of
these light scalars more deeply and provide evidences to decipher $f_0(1500)$
as a primary or fragmented scalar glueball potentially.

\end{abstract}
	
\preprint{\footnotesize  JSNU-PHY-HEP-JUL.25}
\maketitle

	
\newpage
%
%
\section{Introduction}
\label{sect:1}

Recently, the CMS and ALICE experiments located at Large Hadron Collider
reported their measurements about the inner structure of light scalars $f_0(980)$
and $K_0^*(700)$, respectively. They, however, obtained rather different conclusions
about the structure of these light hadrons.
Specifically, the CMS Collaboration claimed that they found strong evidence
of $f_0(980)$ being a normal quark-antiquark state and believed that
`` the results reported in this paper present a clear solution to a half-a-century
puzzle "~\cite{CMS:2023rev}. However, the ALICE Collaboration concluded that their
analyses gave support to $K_0^*(700)$ being a four-quark state, i.e.,
a tetraquark state~\cite{ALICE:2023eyl}. Undoubtedly, these measurements
imply the extraordinarily complicated nature of light scalars discovered half
a century ago. Furthermore, it shows us that understanding
the nature of light scalars clearly is still a challenging task.

In the spectroscopy study~\cite{Workman:2022pdg}, a number of light scalars
are most fascinating objects in the field of strong interactions and are essential
for testing the standard model. It is consensually believed that
light scalars below or near 1 GeV, namely, the isoscalars
$f_0(500)$ (or $\sigma$), $f_0(980)$, the isodoublet $K_{0}^{*}(700)$ (or $\kappa$)
and the isovector $a_0(980)$, form an SU(3) flavor nonet, while those above 1 GeV,
that is, $f_0(1370)$, $f_0(1500)$/$f_0(1710)$, $K_{0}^{*}(1430)$ and $a_0(1450)$,
form another nonet. So far, there are two different scenarios to describe these
scalar mesons under the $q\bar q$ assignment~\cite{Cheng:2005nb}. Explicitly, in
scenario 1 ($S1$), the nonet mesons below 1 GeV are treated as the lowest-lying
states, and those near 1.5 GeV are the first excited states correspondingly. And, in
scenario 2 ($S2$), the nonet mesons near 1.5~GeV are viewed as the $q\bar q$ ground
states, while those below 1 GeV might be the four-quark states. Notwithstanding the
properties of scalar mesons have been intensively studied, the internal structure of
light scalars, especially of $f_0$ (Hereafter, $f_0$ is adopted to denote the isoscalars
$f_0(500, 980, 1370, 1500)$ for simplicity, unless otherwise stated.), is far from
clear understanding yet due to their non-perturbative contents (for a review, see, e.g., Refs.~\cite{Klempt:2007cp,Ochs:2013gi,Chen:2022asf,Gross:2022hyw}).

\begin{figure}[htb]
	\centering
		\includegraphics[scale=0.7]{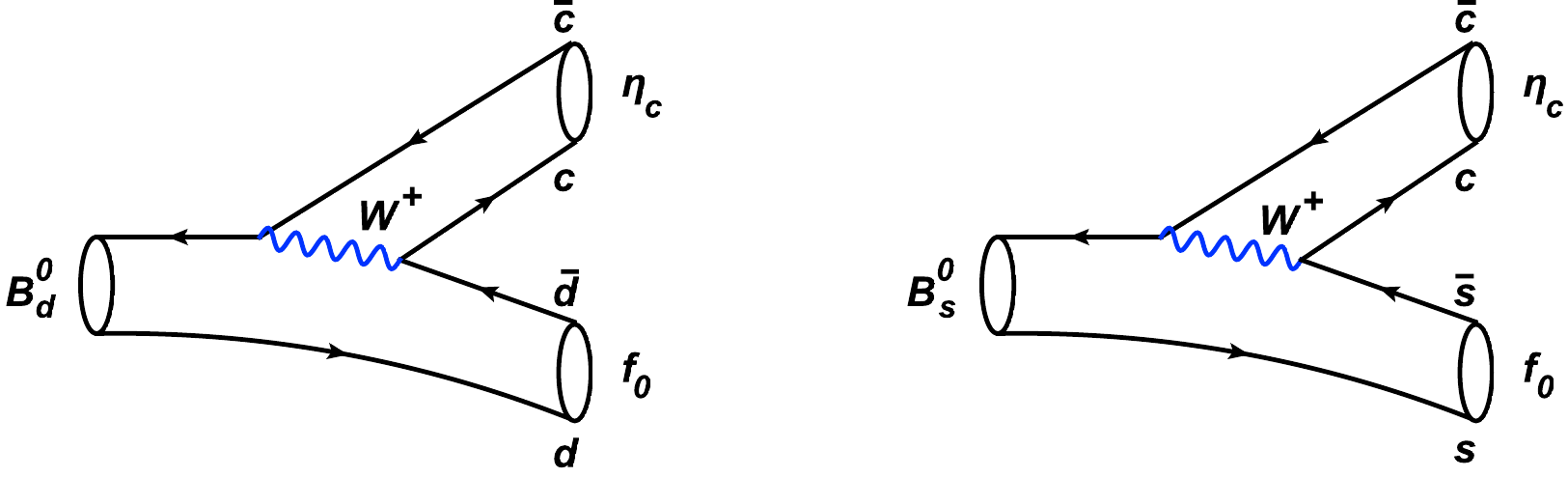}
	\caption{(Color online) Leading quark-level Feynman diagrams
		of $B^0 \to \eta_c f_0$}
	\label{fig:fig1}
\end{figure}

For the isoscalars $f_0$, the relevant investigations would be interesting but also
with challenges because of possible mixing with the scalar glueball~\cite{Cheng:2006hu,
Close:2005vf,Cheng:2015iaa,Li:2021gsx,Li:2024fko}. The related overview on the mixing
of $f_0$ with a primary scalar glueball can be seen in~\cite{Ren:2023ebq},
and references therein. However, rather than primary scalar glueball,
Klempt proposed recently a distinct viewpoint, namely, a fragmented scalar
glueball~\cite{Klempt:2021nuf}. Moreover, Klepmt and Sarantsev estimated the
probabilities of fragmented scalar glueball contributing into different isoscalars
$f_0$~\cite{Klempt:2021wpg}. It is found that the scalar mesons $f_0(1370)$ and
$f_0(1500)$ are contributed from scalar glueball with few percent. Therefore, the
aforementioned isoscalars $f_0$ would mainly be the admixture of light scalar quarkonia
$\frac{u\bar u + d\bar d } {\sqrt{2}}$ and $s\bar s $ with different fractions. In order
to shed light on the nature of $f_0$, even further clarify the involved scalar glueball
contents, the productions of $f_0$ in the $B$-meson decays, e.g., $B$ decaying into
charmonia plus $f_0$, are preferred at both aspects of theory and experiment, since
the phase space in $B$-meson decays is larger than that in $D$-meson decays. Ever since the
$f_0(980)$ meson was reported at the {\it BABAR} and Belle experiments, numerous follow-up
studies were made through $f_0$ productions in the $B$-meson decays to further understand
their nature theoretically and experimentally during the past two decades. Particularly,
as suggested in~\cite{Ochs:2013gi}, the decays like $B^0 \to \eta_c f_0$ could
help differentiate the flavor decomposition of $\frac{u\bar u + d\bar d } {\sqrt{2}}$
and $s\bar s$, as depicted in Fig.~\ref{fig:fig1}. Their branching ratios (BRs) could help
us to determine the ratios of $\frac{u\bar u + d\bar d } {\sqrt{2}}$ and $s\bar s$
correspondingly in the physical $f_0$ states based on the experimental measurements, even
to specify the possible mixing between $\frac{u\bar u + d\bar d } {\sqrt{2}}$ and $s\bar s$.

From the experimental side, the Large Hadron Collider-beauty (LHCb) experiment
ever reported the evidence of $B_s^0 \to \eta_c \pi^+\pi^-$ in 2017, where the
$\pi^+\pi^-$ pair may arise from the decay of resonance $f_0(980)$~\cite{LHCb:2017zar}.
Its BR was measured as, 
\beq
{\rm BR}(B_s^0 \to \eta_c \pi^+ \pi^-) &=&
(1.76 \pm 0.67)\times 10^{-4} \;,
\label{eq:lhcb-e2pi}
\eeq
where various uncertainties have been added in quadrature.
Furthermore, the LHCb detector accomplished its first upgrade and resumed in
2022~\cite{Gao:2024lhc}. It is certainly believed that a huge amount of
superior-quality data can be collected in the near future for precise studies
of heavy quark dynamics and hadron physics. Meantime, the Belle-II experiment
can provide complementary studies to explore the QCD dynamics of $f_0$ and
the related decays.

From the theoretical side, the decays of $B_s^0 \to \eta_c f_0(980)$ and $B_s^0
\to \eta_c f_0(980) (\to \pi^+ \pi^-)$ have been investigated by many groups within
different frameworks. Explicitly,
\begin{itemize}
\item
In 2010, Colangelo, De Fazio, and Wang ever estimated the BRs of $B_s^0 \to \eta_c
f_0(980)$ by employing two different $B_s^0 \to f_0(980)$ form factors~\cite{Colangelo:2010wg}.
The predictions $(4.1 \pm 1.7) \times 10^{-4}$ and $(2.0 \pm 0.9) \times 10^{-4}$ of
${\rm BR}(B_s^0 \to \eta_c f_0(980))$ are generally consistent with each other.

\item
In 2016, Li {\it et al.} studied the quasi-two-body decay of $B_s^0 \to \eta_c f_0(980)
(\to \pi^+ \pi^-)$ with BR being $(3.37^{+1.05}_{-0.77}) \times 10^{-5}$ by employing the
traditional perturbative QCD (PQCD) approach~\cite{Li:2015tja}, which is evidently less
than the LHCb measurement with a factor around 5. Moreover, they also predicted
the BRs of $B_d^0 \to \eta_c f_0(500) (\to \pi^+ \pi^-)$ and $B_s^0 \to \eta_c f_0(1500)
(\to \pi^+ \pi^-)$ as follows,
\beq
{\rm BR}(B_d^0 \to \eta_c f_0(500) (\to \pi^+ \pi^-)) &=&
\biggl\{\begin{array}{ll}
(1.53_{-0.35}^{+0.76})\times 10^{-6} \qquad  \text{(Breit-Wigner model)}
\\
(2.31_{-0.48}^{+0.96})\times 10^{-6}\qquad \text{(Bugg  model)} \end{array} \;,
\\
 {\rm BR}(B_{s}^0 \to \eta_c f_0(1500) (\to \pi^+ \pi^-))
& =& (6.76_{-1.21}^{+1.62})\times 10^{-6}\;.
 \eeq

\item
In 2017, Ke and Li obtained ${\rm BR}(B_s^0 \to \eta_c f_0(980)) = (1.59 \pm 0.57) \times
10^{-4}$~\cite{Ke:2017wni}, combining with $B_s^0 \to f_0(980)$ transition form factor in
the light-front quark model and Wilson coefficient $a_2$ from the measured $B_s^0 \to \eta_c
\phi$ at LHCb experiment. However, it is argued that, if the $\pi^+ \pi^-$ pair comes from
resonance $f_0(980)$ in the measured $B_s^0 \to \eta_c \pi^+ \pi^-$, then this predicted BR
could not support its $q\bar q$ structure.

\item
In 2018, Xie and Li investigated the decays of $B^0$ into $\eta_c$ plus a scalar meson based
on the chiral unitary approach~\cite{Xie:2018rqv}, and got ${\rm BR}(B_s^0 \to \eta_c f_0(980)
(\to \pi^+ \pi^-)) = (1.41 \pm 0.56) \times 10^{-4}$. With the result of $B_s^0 \to \eta_c
\pi^+ \pi^-$, the authors predicted ${\rm BR}(B_d^0 \to \eta_c f_0(500) (\to \pi^+ \pi^-))
= (1.2 \pm 0.5) \times 10^{-5}$. Moreover, the relative ratio between these two BRs was also
given as,
\beq
\frac{{\rm BR}(B_d^0 \to \eta_c f_0(500) (\to \pi^+ \pi^-) )}
{{\rm BR}(B_s^0 \to \eta_c f_0(980) (\to \pi^+ \pi^-))}
&=& (9 \pm 5) \times 10^{-2}\;.
\label{eq:r-xie}
\eeq

\end{itemize}
Evidently, the available predictions are with significant discrepancies and a clear
understanding of QCD dynamics in the light scalars $f_0$ is still tough. Therefore,
further investigations are needed greatly.

To provide more evidences to reveal the internal structure of light scalars,
we shall study the decays $B^0 \to \eta_c f_0$ systematically within the improved
PQCD (iPQCD) framework~\cite{Liu:2018kuo,Liu:2020upy,Liu:2023kxr}.
Here, $B^0$ includes neutral $B_d^0$ and $B_s^0$ mesons,
and the assumptions of quark-antiquark structure for $f_0$ have been made~\footnote{\RyB{\it
As noted in~\cite{Cheng:2005nb}, quantitative predictions based on the four-quark (tetraquark) picture for $f_0(980)$ are challenging, since its decay constant and distribution amplitudes lie beyond the conventional quark model and the relevant nonfactorizable contributions cannot be reliably calculated within current QCD-based factorization approaches. Therefore, in this work, $f_0$ is treated as a two-quark state within the iPQCD framework.}}.
As pointed out in~\cite{Liu:2018kuo,Liu:2020upy,Liu:2023kxr}, the iPQCD formalism
is now self-consistent and ready for $B$-meson decaying into charmonium plus light hadrons
due to the new Sudakov factor that absorbs the charm quark mass effects by $k_T$
resummation, besides of those in hard kernel. The observables such as {\it CP}-averaged
BRs and {\it CP} asymmetries (CPAs) will then be evaluated and analyzed comprehensively.
Because of involving the same $b \to c \bar c s$ transition as the $B_s^0 \to J/\psi
f_0(980)$ mode at quark level~\cite{Liu:2019ymi}, the decays $B_s^0 \to \eta_c f_0$
can provide more information to understand the $B_s^0-\bar B_s^0$ mixing phase $\beta_{s}$
in a complementary manner, which could help us to search for the possible new physics beyond
standard model, as well as to further explore the inner structure of light scalars.

This paper is organized as follows. In Sect.~\ref{sect:2}, we present the formalism and the
perturbative calculations of $B^0 \to \eta_c f_0$. In Sect.~\ref{sect:R&D}, following the
introduction of input parameters, the numerical results and the phenomenological analyses
are presented. Finally, Sect.~\ref{sect:C&S} contains our main conclusions.

%
%
\section{ Formalism and PQCD calculations}
\label{sect:2}

In the standard model, the effective Hamiltonian for the $B^0 \to \eta_c f_0$
decays with $b \to c\bar c q$ transitions, $q$ being light quarks $d$ or $s$,
can be read as~\cite{Buchalla:1995vs}:
\beq
H_{\rm eff}\, &=&\, \frac{G_F}{\sqrt{2}}
\biggl\{ V^*_{cb}V_{cq} \biggl[ C_1(\mu)O_1^{c}(\mu)
+C_2(\mu)O_2^{c}(\mu) \biggr]
- V^*_{tb}V_{tq} \biggl[ \sum_{i=3}^{10}C_i(\mu)O_i(\mu) \biggr] \biggr\}\;,
\label{eq:heff}
\eeq
in which, $G_F$ is the Fermi constant $1.16639\times 10^{-5}{\rm GeV}^{-2}$,
$V_{ij}$ represents the Cabibbo-Kobayashi-Maskawa (CKM) matrix elements,
$C_i(\mu)$ are Wilson coefficients at the renormalization scale $\mu$, and
$O_i(i=1,\cdots,10)$ describe the local four-quark operators,
\begin{enumerate}
	\item[]{(1) Tree operators}
	\begin{eqnarray}
		{\renewcommand\arraystretch{1.5}
			\begin{array}{ll}
				\displaystyle
				O_1^{c}\, =\,
				(\bar{q}_\alpha c_\beta)_{V-A}(\bar{c}_\beta b_\alpha)_{V-A}\;,
				& \displaystyle
				O_2^{c}\, =\, (\bar{q}_\alpha c_\alpha)_{V-A}(\bar{c}_\beta b_\beta)_{V-A}\;,
		\end{array}}
		\label{eq:operators-1}
	\end{eqnarray}
	
	\item[]{(2) QCD penguin operators}
	\begin{eqnarray}
		{\renewcommand\arraystretch{1.5}
			\begin{array}{ll}
				\displaystyle
				O_3\, =\, (\bar{q}_\alpha b_\alpha)_{V-A}\sum_{q'}(\bar{q}'_\beta q'_\beta)_{V-A}\;,
				& \displaystyle
				O_4\, =\, (\bar{q}_\alpha b_\beta)_{V-A}\sum_{q'}(\bar{q}'_\beta q'_\alpha)_{V-A}\;,
				\\
				\displaystyle
				O_5\, =\, (\bar{q}_\alpha b_\alpha)_{V-A}\sum_{q'}(\bar{q}'_\beta q'_\beta)_{V+A}\;,
				& \displaystyle
				O_6\, =\, (\bar{q}_\alpha b_\beta)_{V-A}\sum_{q'}(\bar{q}'_\beta q'_\alpha)_{V+A}\;,
		\end{array}}
		\label{eq:operators-2}
	\end{eqnarray}
	
	\item[]{(3) Electroweak penguin operators}
	\begin{eqnarray}
		{\renewcommand\arraystretch{1.5}
			\begin{array}{ll}
				\displaystyle
				O_7\, =\,
				\frac{3}{2}(\bar{q}_\alpha b_\alpha)_{V-A}\sum_{q'}e_{q'}(\bar{q}'_\beta q'_\beta)_{V+A}\;,
				& \displaystyle
				O_8\, =\,
				\frac{3}{2}(\bar{q}_\alpha b_\beta)_{V-A}\sum_{q'}e_{q'}(\bar{q}'_\beta q'_\alpha)_{V+A}\;,
				\\
				\displaystyle
				O_9\, =\,
				\frac{3}{2}(\bar{q}_\alpha b_\alpha)_{V-A}\sum_{q'}e_{q'}(\bar{q}'_\beta q'_\beta)_{V-A}\;,
				& \displaystyle
				O_{10}\, =\,
				\frac{3}{2}(\bar{q}_\alpha b_\beta)_{V-A}\sum_{q'}e_{q'}(\bar{q}'_\beta q'_\alpha)_{V-A}\;,
		\end{array}}
		\label{eq:operators-3}
	\end{eqnarray}
\end{enumerate}
with $\alpha$, $\beta$ the color indices and $(\bar{q}'q')_{V\pm A} = \bar q'
\gamma_\mu (1\pm \gamma_5)q'$. The index $q'$ in the summation of the above
operators runs through $u,\;d,\;s$, $c$, and $b$. Following Ref.~\cite{Ali:1998eb},
$a_i$ is defined for the standard combination of Wilson coefficients $C_i$ as follows:
\beq
a_{1}&=&C_{2}+\frac{C_{1}}{3}\;,
\qquad
a_{2}=C_{1}+\frac{C_{2}}{3}\;,
\\
a_{i} &=& C_{i}+\frac{C_{i\pm1}}{3}(i=3-10)\;,
\label{eq:wilson-comb}
\eeq
where the upper (lower) sign applies, when $i$ is odd (even).

In the $B^0$-meson rest frame, the momenta $P_{1}$, $P_{2}$ and $P_{3}$ of
$B^0$, $\eta_c$ and $f_0$ correspondingly in light-cone coordinates could be
written as,
\beq
P_{1}&=&\frac{m_{B^0}}{\sqrt{2}}(1, 1, {\bf 0}_{T})\;,
\qquad
P_{2}=\frac{m_{B^0}}{\sqrt{2}}(1-r_{3}^{2}, r_{2}^{2}, {\bf 0}_{T})\;,
\qquad
P_{3}=\frac{m_{B^0}}{\sqrt{2}}(r_{3}^{2}, 1-r_{2}^{2}, {\bf 0}_{T})\;,
\label{eq:coor2}
\eeq
with the ratio $r_{2}=m_{\eta_c}/m_{B^0}$ and $r_{3}=m_{f_{0}}/m_{B^0}$, respectively.
Putting the (light) quark momenta in the $B^0$, $\eta_c$ and $f_{0}$ mesons as $k_{1}$,
$k_{2}$ and $k_{_{3}}$, respectively, then
\beq
k_{1} &=& (x_{1}P_{1}^{+}, 0, {\bf k}_{1T})
 =  (\frac{m_{B^0}}{\sqrt{2}}x_{1}, 0, {\bf k}_{1T})\;,
\label{eq:lqm-b}
\non
k_{2} &=&
(x_{2}P_{2}^{+}, x_{2}P_{2}^{-}, {\bf k}_{2T})
 = ( \frac{m_{B^0}}{\sqrt{2}}x_{2}(1-r_{3}^{2}),  \frac{m_{B^0}}{\sqrt{2}}x_{2} r_{2}^{2}, {\bf k}_{2T})\;,
\label{eq:lqm-mcc}
\\
k_{3} &=&
(x_{3}P_{3}^{+}, x_{3}P_{3}^{-}, {\bf k}_{3T})
 =  (\frac{m_{B^0}}{\sqrt{2}}x_{3}r_{3}^{2}, \frac{m_{B^0}}{\sqrt{2}}x_{3}(1-r_{2}^{2}), {\bf k}_{3T})\;.\nonumber
\label{eq:lqm-fq}
\eeq
Thus the decay amplitude of $B^0 \to \eta_c f_0$
in the iPQCD formalism can be written conceptually as~\cite{Keum:2000wi,Lu:2000em},
\beq
{\cal A}(B^0 \to \eta_c f_0) &\sim &\int\!\! d x_1 d
x_2 d x_3 b_1 d b_1 b_2 d b_2 b_3 d b_3
\non &\times & {\rm Tr}
\left [ C(t) \Phi_{B^0}(x_1, b_1) \Phi_{ \eta_c}(x_2, b_2)
\Phi_{f_0}(x_3, b_3) H(x_i, b_i, t) S_t(x_i)\, e^{-S(t)} \right ]\;.
\label{eq:amp-f1}
\eeq
where $x_{i}$ denotes the fraction of momentum carried by (light) quark
in each meson, $b_{i}$ is the conjugate space coordinate of transverse
momentum $k_{iT}$, $\Phi$ is the wave function to describe the hadronization
of quark-antiquark into a meson. 
$H(x_i, b_i, t)$ is the hard kernel that can be calculated perturbatively,
where $t$ is running scale at the largest energy.
The rest two factors, namely, the Sudakov factor $e^{-S(t)}$ and
the jet function $S_t(x_i)$ as shown in Eq.~(\ref{eq:amp-f1}) above,
play important roles on effective evaluations of the $B$-meson decay amplitude
in the iPQCD approach. Specifically, 
they could kill the end-point singularities~\cite{Botts:1989kf,Botts:1989nd,Li:1992nu}
and smear the double logarithmic divergences~\cite{Li:2001ay,Li:2002mi}. 
The derivation
and the detailed expression of $S_t(x_i)$ and $e^{-S(t)}$ can be easily found in
Refs.~\cite{Botts:1989kf,Botts:1989nd,Li:1992nu,Li:2001ay,Li:2002mi,Li:2003yj,Li:2005kt,
Li:2012nk,Liu:2018kuo,Liu:2020upy,Cheng:2020fcx,Liu:2023kxr}.
For more details and recent advances in the PQCD framework, one can refer to the literature,
for example, see Refs.~\cite{Li:2003yj,Liu:2018kuo,Liu:2020upy,Liu:2023kxr,Cheng:2020fcx}.

\subsection{\boldmath Mixing of $f_0$-mesons}

Before proceeding, a short overview on the mixing of $f_0$-mesons
with their angles, i.e., $\varphi$ of $f_0(500)-f_0(980)$ and
$\varphi'$ of $f_0(1370)-f_0(1500)$, is given essentially.
Generally speaking, based on the assumptions of two-quark structure,
the mixing pattern of $f_0$-meson could be written similarly as
the pseudo-scalar $\eta-\eta'$ mixing, namely,
\beq
\left( \begin{array}{c}
\vert f_0(500) \rangle \\
\vert f_0(980) \rangle \\
\end{array} \right ) &=&
\left( \begin{array}{cc}
	\cos{\varphi} & -\sin{\varphi} \\
	\sin{\varphi} & \; \;\ \cos{\varphi}
\end{array} \right )
\left( \begin{array}{c}
\vert f_{0n} \rangle \\
\vert f_{0s} \rangle \\
\end{array} \right )\;,
\label{eq:mix-ls}
\eeq
and
\beq
\left( \begin{array}{c}
\vert f_0(1370) \rangle \\
\vert f_0(1500) \rangle \\
\end{array} \right ) &=&
\left( \begin{array}{cc}
	\cos{\varphi'} & -\sin{\varphi'} \\
	\sin{\varphi'} & \; \;\ \cos{\varphi'}
\end{array} \right )
\left( \begin{array}{c}
\vert f'_{0n} \rangle \\
\vert f'_{0s} \rangle \\
\end{array} \right )\;,
\label{eq:mix-hs}
\eeq
where $\varphi (\varphi^{\prime})$ is the angle between flavor states
$f_{0n}(f^{\prime}_{0n})$ and $f_{0s} (f^{\prime}_{0s})$ in the $f_0(500)-f_0(980)
[f_0(1370)-f_0(1500)]$ mixing. Notice that, $f_{0n}$ and $f'_{0n}$,
as well as $f_{0s}$ and $f'_{0s}$, have different QCD dynamics, though
they have the same flavor wave function, that is, $f^{(\prime)}_{0n} \equiv
\frac{u\bar u + d\bar d }{\sqrt{2}} $ and $f^{(\prime)}_{0s} \equiv s \bar s$.
It is very clear that $\varphi^{(\prime)}$ could be used as a probe to examine
the deviations from ideal mixing. Also, a clear understanding of
$\varphi^{(\prime)}$ can help us to clarify the structure of the
considered $ f_{0}$.

Unfortunately, due to the currently unknown nature of $f_0(500)$ and $f_0(980)$,
both magnitude and sign of the mixing angle $\varphi$ have not been definitely
determined yet, though various analyses have been made to explore their structure
or mixing angle $\varphi$ from both sides of theory and experiment, e.g., see Refs.~\cite{Li:2012sw,Fritzsch:1972jv,LHCb:2013dkk,Soni:2020sgn,Stone:2013eaa,Liu:2019ymi}.
However, it is noted that, as constrained from the experiments,
$|\varphi| \leq 31^\circ $~\cite{LHCb:2013dkk} and
$|\varphi| \leq 29^\circ $~\cite{Stone:2013eaa} are preferred at $90\%$ confidence level.
Furthermore, $|\varphi| =25^\circ$ was ever deduced phenomenologically through understanding
the available measurements of the decays $B^0 \to J/\psi f_0(500, 980)$ by employing PQCD
approach at the known next-to-leading order (NLO) accuracy~\cite{Liu:2019ymi}.
Moreover, the semi-leptonic decays of $D$-mesons to light scalars provided several
constraints on $\varphi$~\cite{Soni:2020sgn}.

As to the angle $\varphi'$ between the mixing of $f_0(1370)$ and $f_0(1500)$,
there are no more specific discussions currently. Previously, together with
$f_0(1710)$, $f_0(1370)$ and $f_0(1500)$ are usually employed to
decipher the possible scalar glueball from the aspect of hadron spectroscopy.
That is, a general consensus has been achieved that $f_0(1370)$ contains only
a small amount of scalar glueball components and is mainly governed by the
aforementioned $f'_{0n}$. Then, $f_0(1500)$ and $f_0(1710)$ are always considered
as the possible candidates of primary scalar guleball with long-standing controversies,
though different mixing schemes have been established (e.g., see
a short review in Ref.~\cite{Ren:2023ebq} for detail and references therein).
Based on the experimental measurements and the Lattice QCD calculations of related
scalars associated with their decays, two distinct viewpoints were proposed to
clarify that which one of $f_0(1500, 1710)$ is indeed a primary scalar glueball:
Amsler, Close and Zhao {\it et al.} proposed that $f_0(1500)$ is
the primary scalar glueball and $f_0(1710)$ is the scalar meson mixing with glueball
component (e.g., see Refs.~\cite{Close:2000yk,Close:2005vf}), however, Cheng, Chua
and Liu {\it et al.} preferred a very contrary viewpoint (e.g., see Refs.~\cite{Cheng:2006hu,Cheng:2015iaa}).
Recently, an interesting idea named fragmented scalar glueball, rather than primary scalar glueball,
was proposed by Klempt and his collaborator~\cite{Klempt:2021nuf, Klempt:2021wpg}. That is, glueball
is spread over a large number of resonances and the sum of all fractional contributions is close to one.
Then the mixing angle or coefficient of singlet-octet-glueball mixtures in scalars is essential to the
relevant studies containing $f_0(1370, 1500, 1710)$. More interestingly, the analysis performed
in~\cite{Klempt:2021nuf, Klempt:2021wpg} indicates that the scalars $f_0(1370, 1500)$
mix with few scalar-glueball components. Furthermore, the mixing angle $|\phi^s| = (56 \pm 8)^\circ$ is obtained
by neglecting the glueball components in $f_{0}(1370, 1500)$.
Therefore, these two states could be naively considered as the mixtures of flavor states $f_{0n}^\prime$
and $f_{0s}^\prime$ in the quark-flavor basis. Due to different definition of mixing matrix elements
and the relation $\phi^{s}  = \varphi^\prime - 90^\circ$~\cite{Klempt:2021nuf, Klempt:2021wpg},
$\vert \varphi^\prime \vert = (146 \pm 8)^\circ$ is deduced correspondingly for the relevant calculations
in this work. Notice that, the related results could help measure the content of scalar glueball
component mixed in these considered $f_0$ directly.
On the other hand, the future tests on these studies could provide evidence to the controversial issue
mentioned above, i.e., primary scalar glueball $f_0(1500)$ or $f_0(1710)$.

\subsection{\boldmath Wave functions and Distribution amplitudes }

The light-cone wave function of $B^0$ meson in the impact ${\bf b}$ space
can generally be defined as~\cite{Keum:2000wi,Lu:2000em}
\beq
\Phi_{B^0}(x,{\bf b}) &=& \frac{i }{\sqrt{2N_c}} \biggl\{(\psl +m_{B^0}) \gamma_5
\phi_{B^0}(x, {\bf b}) \biggr\}_{\alpha\beta},
\label{eq:wf-B}
\eeq
where the numerically suppressed terms in the PQCD approach has been neglected~\cite{Kurimoto:2001zj,Lu:2002ny,Wei:2002iu}.
In Eq.~(\ref{eq:wf-B}), $\alpha, \beta$ are the Dirac indices, $N_c$ is the color factor, and
$\phi_{B^0}(x, {\bf b})$ is the light-cone distribution amplitude (LCDA)
widely used in the PQCD framework,
\beq
\phi_{B^0}(x, {\bf b})&=& N_{B^0}x^2(1-x)^2
\exp\biggl[-\frac{1}{2}\left(\frac{x m_{B^0}}{\omega_{B^0}}\right)^2
-\frac{\omega_{B^0}^2 {b}^2}{2}\biggr] \;,
\eeq
where $\omega_{B^0}$ is the shape parameter.
In the literature, the shape parameter $\omega_{B^0} = 0.4 (0.5)$ GeV has been fixed
for the $B_d^0 (B_s^0)$ meson through various measurements with good precision~\cite{Kurimoto:2001zj,Lu:2002ny,Wei:2002iu,Ali:2007ff}.
In our calculations, we therefore adopt $\omega_{B^0} = 0.40 \pm 0.04 (0.50
\pm 0.05)$ GeV. $N_{B^0}$ is the normalization constant
obeying the following normalization condition,
\beq
\int_0^1 dx \phi_{B^0}(x, {\bf b}=0) &=& \frac{f_{B^0}}{2 \sqrt{2N_c}}\;.
\label{eq:norm}
\eeq
with the decay constant $f_{B^0}=0.21 (0.23)$ GeV for $B_d^0 (B_s^0)$ meson. 

For the pseudoscalar $\eta_{c}$ meson, its wave function and distribution amplitudes
are taken as those in Ref.~\cite{Chernyak:1983ej,Bondar:2004sv,Chen:2005ht},
\beq
\Phi_{\eta_{c}}(x) &=& \frac{i}{\sqrt{2N_{c}}}\gamma_{5}
\biggl\{\psl \phi_{\eta_{c}}^{v}(x)+m_{\eta_{c}}\phi_{\eta_{c}}^{s}(x)
\biggl\}_{\alpha\beta}\;,
\label{eq:wf-etac}
\eeq
where the twist-2 and twist-3 distribution amplitudes $\phi_{\eta_{c}}^{v}(x)$ and
$\phi_{\eta_{c}}^{s}(x)$ are,
\beq
\phi_{\eta_{c}}^{v}(x) &=& 9.58\frac{f_{\eta_{c}}}{2\sqrt{2N_{c}}}x(1-x)\biggl[\frac{x(1-x)}{1-2.8x(1-x)}\biggl]^{0.7}\;,
\label{eq:da-etac-v}
\\
\phi_{\eta_{c}}^{s}(x) &=& 1.97\frac{f_{\eta_{c}}}{2\sqrt{2N_{c}}}\biggl[\frac{x(1-x)}{1-2.8x(1-x)}\biggl]^{0.7}\;,
\label{eq:da-etac-s}
\eeq
with the decay constant $f_{\eta_c} = 0.39 \pm 0.04$~\cite{Becirevic:2013bsa}
and $x$ being the momentum fraction of charm quark in the $\eta_c$ meson.

For the scalar flavor states $f^{(\prime)}_{0n}$ and $f^{(\prime)}_{0s}$,
following those of the light scalars evaluated in QCD sum rules, their
light-cone wave functions can then be written as~\cite{Cheng:2005ye,Li:2008tk},
\beq
\Phi_{f_{0}}(x) &=& \frac{i}{\sqrt{2N_{c}}}
\biggl\{\psl\phi_{f_{0}}(x)+m_{f_{0}}\phi_{f_{0}}^S(x)+m_{f_{0}}(\nsl\vsl-1)
\phi_{f_{0}}^T(x)\biggl\}_{\alpha\beta}\;,
\label{eq:wf-f0}
\eeq
with the twist-2 and twist-3 LCDAs $\phi_{f_{0}}(x)$ and $\phi_{f_{0}}^{S,T}(x)$,
respectively. These LCDAs can be expanded as the Gegenbauer polynomials in the
following form~\cite{Lu:2006fr}:
\beq
\phi_{f_{0}}(x) &=& \frac{\bar{f}_{f_{0}}(\mu)}{2\sqrt{2N_{c}}}
\biggl\{6x(1-x)\sum_{m=1}^{\infty}B_m(\mu)C_m^{3/2}(2x-1)\biggl\}\;,
\label{eq:lcda-f0}
\\
\phi_{f_{0}}^S(x) &=& \frac{\bar{f}_{f_{0}}(\mu)}{2\sqrt{2N_{c}}}
\biggl\{1+\sum_{m=1}^{\infty}a_m(\mu)C_m^{1/2}(2x-1)\biggl\}\;,
\label{eq:lcda-s-f0-G}
\\
\phi_{f_{0}}^T(x) &=& \frac{\bar{f}_{f_{0}}(\mu)}{2\sqrt{2N_{c}}}\frac{d}{dx}\biggl\{x(1-x)\times \biggl[1+\sum_{m=1}^{\infty}b_m(\mu)C_m^{3/2}(2x-1)\biggl]\biggl\},
\label{eq:lcda-t-f0-G}
\eeq
where $\bar f_{f_0}(\mu)$ stands for the scalar decay constant of light scalars $f_0$ at the renormalization scale $\mu$,
$B_m(\mu)$, $a_m(\mu)$ and $b_m(\mu)$ are the Gegenbauer moments and $C_m^{3/2}$
and $C_m^{1/2}$ are the Gegenbauer polynomials. Due to the absent Gegenbauer moments
$a_m(\mu)$ and $b_m(\mu)$ in $S1$, we have to adopt their asymptotic forms presently
for simplicity in our calculations, that is~\cite{Cheng:2005ye},
\beq
\phi_{f_{0}}^S(x) &=& \frac{1}{2\sqrt{2N_{c}}}\bar{f}_{f_{0}}(\mu)\;,
\label{eq:lcda-s-f0-A}
\\
\phi_{f_{0}}^T(x) &=& \frac{1}{2\sqrt{2N_{c}}}\bar{f}_{f_{0}}(\mu)(1-2x).
\label{eq:lcda-t-f0-A}
\eeq

\subsection{\boldmath Perturbative Calculations of $B^0 \to \eta_c f_0$}

It is well known that, compared to QCD factorization approach~\cite{Beneke:1999br} and
soft-collinear effective theory~\cite{Bauer:2004tj} based on the collinear
factorization theorem, the PQCD approach can be
used safely to calculate the nonfactorizable emission ($nfe$)
diagrams and the annihilation ones, besides the factorizable
emission ($fe$) contributions. In the PQCD framework, the divergences at the end-points can be
eliminated by keeping transverse momentum of the valence quark,
and the resultant Sudakov factor makes it more self-consistent.
In particular, the newly derived Sudakov factor for $c\bar c$-meson
by including the charm quark mass effects further
improves the PQCD framework for the decays of $B$-meson into charmonium
plus light hadron(s)~\cite{Liu:2018kuo,Liu:2020upy}.

\begin{figure}[!!htb]
	\begin{center}
		\includegraphics[scale=0.8]{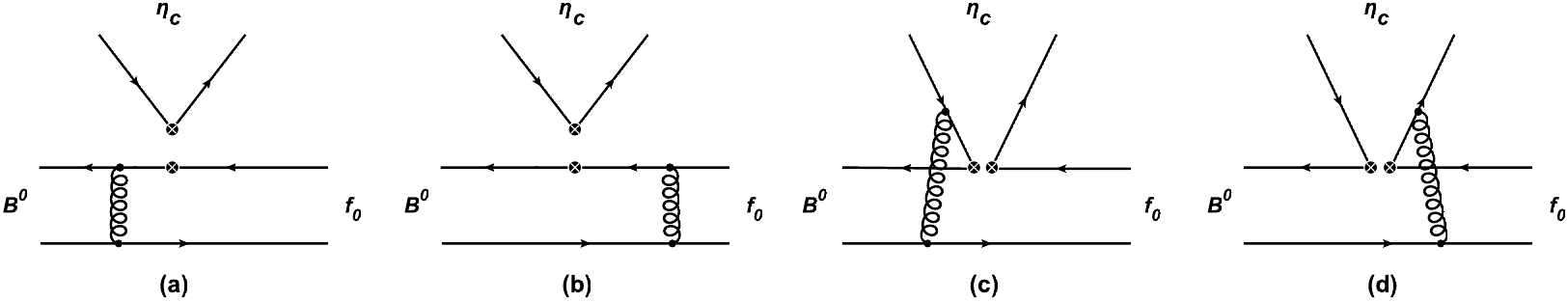}
		\caption{
        Typical Feynman diagrams contributing to the $B^{0} \to \eta_c f_0$
        decays in the PQCD approach at leading order.}
		\label{fig:fig2}
	\end{center}
\end{figure}

As depicted in Fig.~\ref{fig:fig2}, the first two diagrams Fig.~\ref{fig:fig2}(a)
and~\ref{fig:fig2}(b) will contribute the leading order (LO) factorizable emission
amplitudes $F_{fe}$, while the latter two Fig.~\ref{fig:fig2}(c)
and~\ref{fig:fig2}(d) will provide the LO non-factorizable emission ones $M_{nfe}$.
In the PQCD calculations of $B^0 \to \eta_c f_0$, 
the superscripts $LL, LR, SP$ stand for the contributions induced by
$(V - A)(V - A)$, $(V - A)(V + A)$ and $(S - P)(S + P)$ operators, respectively.
It is emphasized that the $(S - P)(S + P)$ operators come from Fierz transformation
of the $(V - A)(V + A)$ ones.

Then, the relevant factorization formulas could be
presented explicitly as follows,
\beq
F_{fe}^{LL} &=&
8 \pi C_{F} f_{\eta_{c}} m^{4}_{B^0} \int^{1}_{0} d x_1 d x_3
\int^{\infty}_{0} b_1 d b_1  b_3d b_3 \phi_{B^0}(x_1,b_1)
\non &&
\times \Biggl\{\Biggl[r_{3} (2(r_{2}^{2}-1)x_{3}+r_{2}^{2}+1) \phi_{f_0}^{S}(x_3)
+r_{3}(r_{2}^{2}-1)(2x_{3}-1)\phi_{f_0}^{T}(x_3)
+((1-r_{2}^{2})+x_{3} (1-2r_{2}^{2}))\phi_{f_0}(x_3)\Biggr]
\non &&
\cdot h_{fe}(x_{1},x_{3},b_{1},b_{3})E_{fe}(t_{a})
+ \Biggl[2(1-r_{2}^{2})r_{3}\phi_{f_0}^{S}(x_3)\Biggr]
h_{fe}(x_{3},x_{1},b_{3},b_{1})E_{fe}(t_{b})
\Biggr\}  \;,
\label{eq:DecAmp-fe-LL}
\eeq
with the color factor $ C_{F} = {4 \over 3}$.
\beq
F_{fe}^{LR}&=&-F_{fe}^{LL} \;,
\label{eq:DecAmp-fe-LR}
\eeq
\beq
M_{nfe}^{LL} &=& 	-\frac{32 \pi C_{F} m^{4}_{B^0}}{\sqrt{6}}
\int^{1}_{0} d x_1 d x_2 d x_3 \int^{\infty}_{0} b_1 d b_1 b_2d b_2
\phi_{B^0}(x_1,b_1) \phi_{\eta_c}^{v}(x_2)
	\non &&
\times
\Biggl\{\Biggl[x_3  (\phi_{f_0}(x_3) -2 r_{2}^{2} \phi_{f_0}(x_3)
+ 2r_{3}(r_{2}^{2} - 1 ) \phi_{f_0}^{T}(x_3)) \Biggr]
h_{nfe}(x_{1},x_{2},x_{3},b_{1},b_{2})E_{nfe}(t_{nfe}) \Biggr\}.
	\label{eq:DecAmp-nfe-LL}
\eeq
\beq
	M_{nfe}^{SP} &=& M_{nfe}^{LL} \;,
	\label{eq:DecAmp-nfe-SP}
\eeq
In the above factorization formulas, the explicit forms of hard functions
$h_{fe, nfe}(x_i, b_i)$ and evolution functions $E_{fe, nfe}(t)$ could be
found easily in the literature with PQCD approach, e.g.,
Refs.~\cite{Liu:2013nea,Liu:2019ymi,Xiao:2019mpm}.

\begin{figure}[!!htb]
	\begin{center}
		\includegraphics[scale=0.8]{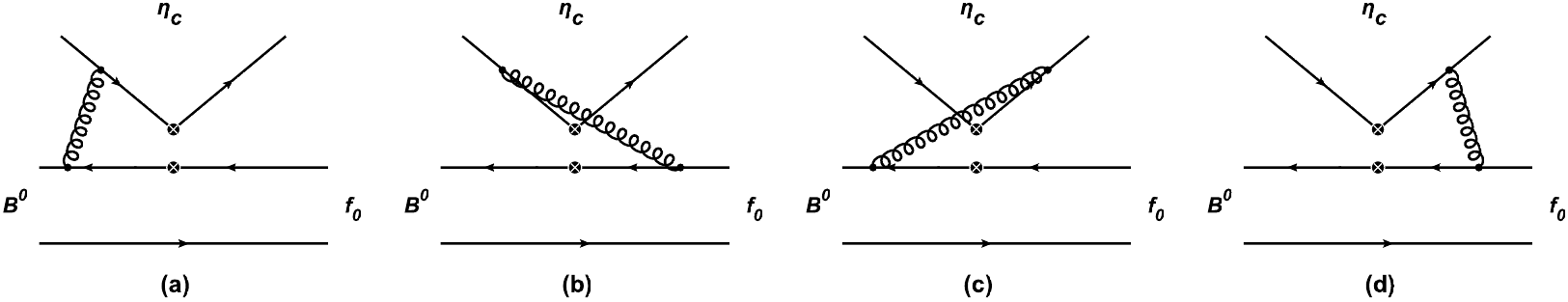}
		\caption{
        Vertex corrections to the $B^{0} \to \eta_c f_0$ decays.}
		\label{fig:fig3}
	\end{center}
\end{figure}

As stated in Refs.~\cite{Liu:2013nea,Liu:2009yno,Liu:2019ymi,Xiao:2019mpm},
the $B$-meson decays into a charmonium plus light hadron(s)
are color-suppressed. However, the significant NLO contributions such as vertex
corrections (as illustrated in Fig.~\ref{fig:fig3}) and NLO Wilson coefficients
can effectively improve the numerical results of BRs and make the predictions
close to, even consistent with the experimental data within uncertainties.
Therefore, these contributions should be naturally included in the investigations
of $B^0 \to \eta_c f_0$.

By considering various contributions from Figs.~\ref{fig:fig2} and \ref{fig:fig3},
the decay amplitudes of $B^{0}_{d} \to \eta_c f_{0n}^{(\prime)}$ and $B^{0}_{s}
\to \eta_c f_{0s}^{(\prime)}$ could be directly written as
\beq
\xi A(B_{d(s)}^{0} \to \eta_{c}\ f_{0n(s)}^{(\prime)})
&=&
F_{fe}^{LL}
\biggl\{ V_{cb}^{\ast} V_{cd(s)}
\tilde{a}_{2}-V_{tb}^{\ast} V_{td(s)}
\biggl( \tilde{a}_{3}- \tilde{a}_{5}- \tilde{a}_{7}+ \tilde{a}_{9}\biggr) \biggr\}
\non &&
+ M_{nfe}^{LL} \biggl\{V_{cb}^{\ast} V_{cd(s)} C_{2} -V_{tb}^{\ast} V_{td(s)}
\biggl( C_{4}+C_{6}+C_{8}+C_{10}\biggr) \biggr\}\;,
\label{eq:DecAmp-etac}
\eeq
where $\xi$ is $\sqrt{2}$ and $1$ for the final state $f_{0n}^{(\prime)}$
and $f_{0s}^{(\prime)}$, respectively. Notice that, here, the related
vertex corrections have been included in the effective Wilson coefficients
$\tilde{a}_{i}$, whose expressions can be found easily in the literature,
for example, see Refs.~\cite{Song:2002gw,Xiao:2019mpm}.

An essential comment on the Wilson coefficients is in order. As stated in
Ref.~\cite{Liu:2013nea}, the PQCD calculations at LO use the LO Wilson
coefficients $C_{i}(m_{W})$ and the LO renormalization group evolution
matrix $U(t,m)^{(0)}$ for the Wilson coefficient associated with the LO
running coupling $\alpha_{s}$,
\beq
\alpha_{s}(t)&=&\frac{4\pi}{\beta_{0}\ln [t^{2}/\Lambda_{\rm QCD}^{2}]},
\eeq
where $\beta_{0}=(33-2N_{f})/3$ with $N_{f}=4$ or $N_{f}=5$ corresponding
to $t<m_{b}$ or $t\geq m_{b}$. Considering the PQCD calculation at NLO
accuracy, it is natural to include the NLO Wilson coefficients $C_{i}(m_{W})$
and the NLO renormalization group evolution matrix $U(t,m,\alpha)$
( The interested readers can see the Eq.~(7.22) in Ref.~\cite{Buchalla:1995vs}
for detail. ) with the running coupling $\alpha_{s}(t)$ at two-loop,
\beq
\alpha_{s}(t)&=&\frac{4\pi}{\beta_{0}\ln(t^{2}/\Lambda_{\rm QCD}^{2})}\cdot \biggl
\{1-\frac{\beta_{1}}
{\beta_{0}^{2}}\cdot \frac{\ln
	[\ln(t^{2}/\Lambda_{\rm QCD}^{2})]}{\ln(t^{2}/\Lambda_{\rm QCD}^{2})}\biggr\},
\eeq
where $\beta_{1}=(306-38N_{f})/3$. For the hadronic scale $\Lambda_{\rm QCD}$, the
$\Lambda_{\rm QCD}^{(4)}=0.287$ GeV (0.326 GeV) could be got by using
$\Lambda_{\rm QCD}^{(5)}=0.225$ GeV for the LO (NLO) case~\cite{Buchalla:1995vs}.
For the hard scale $t$, the cut-off $\mu_{0}=1.0$ GeV is chosen~\cite{Xiao:2008sw}.

Combining the above amplitudes in Eq.~(\ref{eq:DecAmp-etac}) and
quark-flavor mixing schemes in Eqs.~(\ref{eq:mix-ls}) and~(\ref{eq:mix-hs}),
the decay amplitudes of the decays $B^0 \to \eta_c f_0(500, 980)$
could be given as,
\beq
	{\cal A}(B_{d}^{0}\to \eta_c f_{0}(500))
	&=& A(B_{d}^{0}\to \eta_c f_{0n}) \cos \varphi \;,
	\label{eq:DecAmp-etacf05-d}
\qquad
	{\cal A}(B_{d}^{0}\to \eta_c f_{0}(980))
	= A(B_{d}^{0}\to \eta_c f_{0n}) \sin \varphi \;,
	\label{eq:DecAmp-etacf09-d}
\\
	{\cal A}(B_{s}^{0}\to \eta_c f_{0}(500))
	&=& - A(B_{s}^{0}\to \eta_c f_{0s}) \sin \varphi \;,
	\label{eq:DecAmp-etacf05-s}
\qquad
	{\cal A}(B_{s}^{0}\to \eta_c f_{0}(980))
	= A(B_{s}^{0}\to \eta_c f_{0s}) \cos \varphi \;.
	\label{eq:DecAmp-etacf09-s}
\eeq
Similarly, for the decays $B^0 \to \eta_c f_0(1370, 1500)$, we have
\beq
	{\cal A}(B_{d}^{0}\to \eta_c f_{0}(1370))
	&=& A(B_{d}^{0}\to \eta_c f_{0n}^\prime) \cos \varphi' \;,
	\label{eq:DecAmp-etacf13-d}
\quad
	{\cal A}(B_{d}^{0}\to \eta_c f_{0}(1500))
	= A(B_{d}^{0}\to \eta_c f_{0n}^\prime) \sin \varphi' \;.
	\label{eq:DecAmp-etacf15-d}
\\
	{\cal A}(B_{s}^{0}\to \eta_c f_{0}(1370))
	&=& - A(B_{s}^{0}\to \eta_c f_{0s}^\prime) \sin \varphi' \;,
	\label{eq:DecAmp-etacf13-s}
\quad
	{\cal A}(B_{s}^{0}\to \eta_c f_{0}(1500))
	= A(B_{s}^{0}\to \eta_c f_{0s}^\prime) \cos \varphi' \;.
	\label{eq:DecAmp-etacf15-s}
\eeq
Then, the $B^0 \to \eta_c f_0$ BR could be written as
\beq
{\rm BR}(B^0 \to \eta_c f_0) &\equiv&
\tau_{B^{0}} \Gamma(B^{0} \to \eta_c f_0)
= \tau_{B^0} \cdot
\frac{G_F^2}{32\pi m_{B^0}} \cdot \Phi(r_2, r_3)
\cdot |{\cal A}(B^0 \to \eta_c f_0)|^2\;.
\label{eq:BR}
\eeq
where $\tau_{B^0}$ is the lifetime of neutral $B$-meson and
$\Phi(r_2, r_3)$ stands for the phase space factor of $B^{0}
\to \eta_c f_0$ with $ \Phi(x,y) \equiv \sqrt{[1-(x-y)^2]
[1-(x+y)^2]}$~\cite{Fleischer:2011au}.

%
%
\section{Numerical results and discussions}
\label{sect:R&D}
	
In this section, we present theoretical predictions about BRs and CPAs
of the $B^{0} \to \eta_c f_0$ decays. In the numerical calculations,
several input parameters are collected and introduced in order.
\begin{itemize}
	\item
Meson masses (GeV), decay constants (GeV) and lifetimes (ps)~\cite{Workman:2022pdg},
\beq
m_W &=& 80.40,\qquad
m_{B_d^0}= 5.28,\qquad
m_{B_s^0}= 5.37,\qquad
m_b = 4.8, \qquad
m_c = 1.5, \non
\tau_{B_d^0} &=& 1.519,\qquad
\tau_{B_s^0} = 1.520,\qquad
m_{\eta_{c}} = 2.98,\qquad
m_{f_{0}(500)}=0.513,
\\
m_{f_{0}(980)} &=& 0.980, \qquad
m_{f_{0}(1370)} = 1.350, \qquad
m_{f_{0}(1500)} = 1.506.
\nonumber
\label{eq:inputs}
\eeq
The quark-flavor state masses $m_{f_{0n}}$ and $m_{f_{0s}}$
can be calculated through the following relation~\cite{Alford:2000mm,Cheng:2002ai,Anisovich:2001zp,Anisovich:2002wy,Anisovich:2000uh,Gokalp:2004ny,Maiani:2004uc,Kleefeld:2001ds,Verma:2011yw},
\beq
m_{f_{0n}}^2 &=& m_{f_0(500)}^2 \cos^2\varphi + m_{f_0(980)}^2 \sin^2\varphi\;,
\label{eq:mass-fn}
\quad
m_{f_{0s}}^2 = m_{f_0(500)}^2 \sin^2\varphi + m_{f_0(980)}^2 \cos^2\varphi\;.
\label{eq:mass-fs}
\eeq
With the corresponding replacements of $f_0(500) \to f_0(1370)$, $f_0(980)
\to f_0(1500)$, and $\varphi \to \varphi'$, one could obtain the other set
of quark-flavor state masses $m_{f_{0n}^\prime}$ and $m_{f_{0s}^\prime}$
easily.

\item
Following Refs.~\cite{Ren:2023ebq,Liu:2019ymi,Cheng:2005ye}, the scalar decay
constants and Gegenbauer moments at renormalization scale $\mu =1$~GeV in the
LCDAs of $f_{0n}^{(\prime)}$ and $f_{0s}^{(\prime)}$ are chosen as,
\beq
\bar f_{f_{0n}} &\simeq& 0.35 \;,
\qquad
B_1^n = -0.92\pm 0.08 \;,
\qquad
B_3^n = -1.00\pm 0.05 \;,
\\
\bar f_{f_{0s}} &\simeq& 0.33 \;,
\qquad
B_{1,3}^s \simeq 0.8 B_{1,3}^n\;.
\eeq
and
\beq
\bar f_{f_{0n}^\prime} &=& \biggl\{\begin{array}{cc}
-0.280\pm 0.030 \qquad (S1)
\\
\hspace{3mm}
0.460\pm 0.050 \qquad (S2)
\end{array} \;,
\qquad
\bar f_{f_{0s}^\prime} = \biggl\{\begin{array}{cc}
-0.255\pm 0.030 \qquad (S1)
\\
\hspace{3mm}
0.490\pm 0.050 \qquad  (S2)
\end{array}\;,
\\
B_1^{n^\prime} &=& \biggl\{\begin{array}{cc}
\hspace{3mm}
1.00 \pm 0.50 \quad (S1)
\\
-0.60\pm 0.14 \quad (S2)
\end{array} \;,
\quad
B_3^{n^\prime} = \biggl\{\begin{array}{cc}
-1.65\pm 0.18 \quad (S1)
\\
-0.46\pm 0.25 \quad (S2)
\end{array} \;,
\quad
B_{1,3}^{s^\prime}  = 0.8 B_{1,3}^{n^\prime}\;.
\eeq

\item
For the CKM matrix elements, we adopt the Wolfenstein parametrization
up to $\mathcal{O}(\lambda^{6})$~\cite{Wolfenstein:1983yz,Descotes-Genon:2017thz}
and the updated parameters~\cite{Workman:2022pdg}: $A=0.826$, $\lambda=0.2250$,
$\bar\rho=0.159\pm 0.010$, $\bar\eta=0.348\pm 0.010$.
\end{itemize}

\begin{table}[htb]
	\caption{BRs of $B^0 \to \eta_c f_0(500, 980)$ in iPQCD approach}
	\label{tab:BR-ls}
	\begin{center}\vspace{-0.5cm}{
			\begin{tabular}[t]{c| c}
				\hline \hline
				Decay modes   &  Branching ratios   \\
				\hline
				$B_d^0 \to \eta_c f_0(500)$
				&$5.10^{+1.46}_{-1.10}(\omega_{B^0})
				^{+1.10}_{-1.49}(f_{\eta_c})
                ^{+1.07}_{-0.97}(\bar{f}_{f_{0n}})
				^{+0.40}_{-0.39}(B_{i}^n)
				^{+0.14}_{-0.33}(a_t)
				^{+0.20}_{-0.22}(\varphi)
				\times 10^{-5}$\;
				\\
				$B_d^0 \to \eta_c f_0(980)$
				&$1.03^{+0.29}_{-0.20}(\omega_{B^0})
				^{+0.22}_{-0.30}(f_{\eta_c})
                ^{+0.21}_{-0.20}(\bar{f}_{f_{0n}})
				^{+0.08}_{-0.08}(B_{i}^n)
				^{+0.03}_{-0.07}(a_t)
				^{+0.20}_{-0.19}(\varphi)
				\times 10^{-5}$\;			
				\\
				$B_s^0 \to \eta_c f_0(500)$
				&$1.49^{+0.48}_{-0.36}(\omega_{B^0})
				^{+0.32}_{-0.44}(f_{\eta_c})
                ^{+0.31}_{-0.29}(\bar{f}_{f_{0s}})
				^{+0.12}_{-0.13}(B_{i}^s)
                ^{+0.05}_{-0.09}(a_t)
				^{+0.29}_{-0.27}(\varphi)
				\times 10^{-4}$\;			
				\\
				$B_s^0 \to \eta_c f_0(980)$
				&$6.37^{+2.08}_{-1.51}(\omega_{B^0})
				^{+1.37}_{-1.86}(f_{\eta_c})
                ^{+1.33}_{-1.21}(\bar{f}_{f_{0s}})
				^{+0.53}_{-0.52}(B_{i}^s)
				^{+0.24}_{-0.39}(a_t)
				^{+0.25}_{-0.27}(\varphi)
				\times 10^{-4}$\;
				\\
				\hline \hline
		\end{tabular}}
	\end{center}
\end{table}	

Now, by employing the decay amplitudes shown in
Eqs.~(\ref{eq:DecAmp-etacf05-d})-(\ref{eq:DecAmp-etacf15-s})
and~(\ref{eq:BR}), we can calculate the BRs of
$B^{0} \to \eta_c f_0(500, 980)$ in the iPQCD formalism at NLO level. The numerical
results are presented in Table~\ref{tab:BR-ls} explicitly. The theoretical
uncertainties are induced sequentially by the shape parameter $\omega_{B^{0}}$,
the decay constant $f_{\eta_c}$, the scalar decay constants $\bar f_{f_{0n}}$
and $\bar f_{f_{0s}}$, the Gegenbauer moments $B_1^{n,s}$ and $B_3^{n,s}$,
the factor $a_t = 1.0 \pm 0.2$ of hard scale $t_{\rm max}$
describing the effects from remaining higher order QCD contributions,
and the mixing angle $|\varphi| = (25.0\pm 2.5)^\circ$, respectively.
By the way, the uncertainties from the CKM parameters $V(\bar \rho, \bar \eta)$
are too small to be taken into account of BRs. It is clear that the hadronic
parameters such as $\omega_{B^0}$, 
$f_{\eta_c}$ and $\bar f_{f_0}$ play important roles in iPQCD BRs.
Therefore, in order to improve our theoretical predictions promisingly,
these nonperturbative parameters further constrained by
the Lattice QCD calculations and/or the increasingly precise data
are demanded urgently.

From Table~\ref{tab:BR-ls}, it is evident that ${\rm BR}(B_s^0 \to \eta_c f_0(980))
= (6.37^{+2.90}_{-2.77}) \times 10^{-4}$ predicted in the iPQCD formalism is consistent
with those in Ref.~\cite{Colangelo:2010wg} within still large uncertainties.
Although the decays $B^{0} \to \eta_c f_0(500, 980)$ have never been observed by any
experiments, their large BRs around $10^{-5} \sim 10^{-4}$ are measurable and could
be tested at LHCb and Belle-II experiments in the near future. And, in principle,
their measurements could help hint the mixing angle $\varphi$
between $f_0(500)$ and $f_0(980)$ under the $q\bar q$ assignment.

In light of the large uncertainties induced by nonperturbative QCD parameters,
we shall define several interesting ratios between the BRs of the decays
$B^0 \to \eta_c f_0(500, 980)$ on the basis of the iPQCD values shown in
Table~\ref{tab:BR-ls}. In these ratios, it is expected that theoretical
uncertainties arising from the same parameter might be greatly canceled
in the predictions of many observables. Specifically,
\begin{itemize}
\item
Involving mixing angle only,
\beq
R_{d}[f_0(500)/f_0(980)] &\equiv&
	\frac{{\rm BR}(B_{d}^{0} \to \eta_c f_{0}(500))}
	{{\rm BR}(B_{d}^{0} \to \eta_c f_{0}(980))}
	= \frac{\Phi^{d}_{f_0(500)}}{\Phi^{d}_{f_0(980)}}\cdot \cot^2 \varphi
	\approx
	4.95^{+0.86}_{-0.65}
    \;,  \label{eq:r59d}
    \\
R_{s}[f_0(980)/f_0(500)] &\equiv&
		\frac{{\rm BR}(B_{s}^{0} \to \eta_c f_{0}(980))}
		{{\rm BR}(B_{s}^{0} \to \eta_c f_{0}(500))}
		= \frac{\Phi^{s}_{f_0(980)}}{\Phi^{s}_{f_0(500)}}\cdot \cot^2 \varphi
		\approx
		4.28^{+0.72}_{-0.56}
        \;, \label{eq:r95s}
		\eeq
where $\Phi^{d}_{f_0(500)}/\Phi^{d}_{f_0(980)} \approx 1.083$ and $\Phi^{s}_{f_0(980)}
/ \Phi^{s}_{f_0(500)} \approx 1.077$. It is clear that the above two ratios involving
only $\cot^2\varphi$ are clean with no significant pollution by various nonperturbative
parameters. The underlying reason is that these two ratios contain the same transition
amplitudes $B_d^0 \to f_{0n}$ and $B_s^0 \to f_{0s}$, respectively. In principle, they
can provide the best ways to determine the magnitude of mixing angle $\varphi$ promisingly.

\item
Involving SU(3) flavor symmetry-breaking effects only,
\beq
R_{d}^{s}[f_0(980)/f_0(500)] &\equiv&
\frac{{\rm BR}(B_{s}^{0} \to \eta_c f_{0}(980))}
{{\rm BR}(B_{d}^{0} \to \eta_c f_{0}(500))}
\non
&=& \frac{\tau_{B_s^0}}{\tau_{B_d^0}}\cdot (\frac{m_{B_s^0}}{m_{B_d^0}})^7
\cdot \frac{\Phi^{s}_{f_0(980)}}{\Phi^{d}_{f_0(500)}}
\cdot \frac{|{\cal A}(B_s^0 \to \eta_c f_{0s})/m^2_{B_s^0}|^2}{|{\cal A}(B_d^0 \to \eta_c f_{0n})/m^2_{B_d^0}|^2}
\approx
12.49^{+0.41}_{-0.35}\;,\label{eq:rs9d5}
\\
R_{d}^{s}[f_0(500)/f_0(980)] &\equiv&
\frac{{\rm BR}(B_{s}^{0} \to \eta_c f_{0}(500))}
{{\rm BR}(B_{d}^{0} \to \eta_c f_{0}(980))}
\non
&=& \frac{\tau_{B_s^0}}{\tau_{B_d^0}}\cdot (\frac{m_{B_s^0}}{m_{B_d^0}})^7
\cdot \frac{\Phi^{s}_{f_0(500)}}{\Phi^{d}_{f_0(980)}}
\cdot \frac{|{\cal A}(B_s^0 \to \eta_c f_{0s})/m^2_{B_s^0}|^2}{|{\cal A}(B_d^0 \to \eta_c f_{0n})/m^2_{B_d^0}|^2}
\approx
14.47^{+0.47}_{-0.87}
\;.\label{eq:rs5d9}
\eeq
Because the values of lifetimes, masses, and phase-space factors in the $B_s^0$
and $B_d^0$ decay channels are definite, then these two ratios in Eqs.~(\ref{eq:rs9d5})
and~(\ref{eq:rs5d9}) with very small theoretical errors can be used to measure the SU(3)
flavor symmetry-breaking effects in between the decay amplitudes of $B_d^0 \to \eta_c f_{0n}$
and $B_s^0 \to \eta_c f_{0s}$.

\item
Entanglement of mixing angle and broken SU(3) flavor symmetry,
\beq
R_{d}^{s}[f_0(500)] &\equiv&
\frac{{\rm BR}(B_{s}^{0} \to \eta_c f_{0}(500))}
{{\rm BR}(B_{d}^{0} \to \eta_c f_{0}(500))}
\non
&=& \frac{\tau_{B_s^0}}{\tau_{B_d^0}}\cdot (\frac{m_{B_s^0}}{m_{B_d^0}})^7
\cdot \frac{\Phi^{s}_{f_0(500)}}{\Phi^{d}_{f_0(500)}}
\cdot \frac{|{\cal A}(B_s^0 \to \eta_c f_{0s})/m^2_{B_s^0}|^2}
{|{\cal A}(B_d^0 \to \eta_c f_{0n})/m^2_{B_d^0}|^2}\cdot \tan^2 \varphi
\approx
 2.92^{+0.45}_{-0.43}
\;,\label{eq:rsd5}
\\
R_{d}^{s}[f_0(980)] &\equiv&
\frac{{\rm BR}(B_{s}^{0} \to \eta_c f_{0}(980))}
{{\rm BR}(B_{d}^{0} \to \eta_c f_{0}(980))}
\non
&=& \frac{\tau_{B_s^0}}{\tau_{B_d^0}}\cdot (\frac{m_{B_s^0}}{m_{B_d^0}})^7
\cdot \frac{\Phi^{s}_{f_0(980)}}{\Phi^{d}_{f_0(980)}}
\cdot \frac{|{\cal A}(B_s^0 \to \eta_c f_{0s})/m^2_{B_s^0}|^2}
{|{\cal A}(B_d^0 \to \eta_c f_{0n})/m^2_{B_d^0}|^2}\cdot \cot^2 \varphi
\approx	
61.84^{+11.02}_{-8.67}
\;.\label{eq:rsd9}
\eeq
It is noteworthy that these two ratios are a bit complicated
because of the entanglement of broken SU(3) flavor symmetry
and mixing angle. However, they could provide constraints
supplementarily to the mixing angle and the broken SU(3)
flavor symmetries. For example, if $\varphi$ could be
determined through Eqs.~(\ref{eq:r59d}) and~(\ref{eq:r95s})
definitely, then the flavor symmetry-breaking effects in
these decays could be deduced, and vice versa.
\end{itemize}
It is emphasized that the uncertainties of ratios $R_d[f_0(500)/f_0(980)]$,
$R_s[f_0(980)/f_0(500)]$, $R^s_{d}[f_0(500)]$, and $R^s_{d}[f_0(980)]$ are mainly
from the variation of $\varphi$. Generally speaking, the uncertainties induced by
hadronic parameters are canceled to a great extent in the above interesting ratios,
e.g., see Eqs.~(\ref{eq:rs9d5}) and~(\ref{eq:rs5d9}). These clean ratios with large
values could be tested in the near future experiments. Furthermore, the future
experimental confirmations on these predictions might provide evidences to support
the CMS measurement of $f_0(980)$ introduced in Sect.~\ref{sect:1}.

To compare with the available data mentioned in Sect.~\ref{sect:1}, we further analyze
the decay $B_s^0 \to \eta_c f_0(980)$ under the narrow-width approximation (NWA) with
branching fraction ${\cal B}(f_0(980) \to \pi^+ \pi^- ) = 0.45^{+0.07}_{-0.05}$~\cite{Liu:2019ymi}.
The $B^0 \to \eta_c \pi^+ \pi^-$ BR via the resonance state $f_0(980)$
could then be written as,
\beq
{\rm BR}(B^0 \to \eta_c \pi^+ \pi^-) &\equiv& {\rm BR}(B^0 \to \eta_c f_0(980)) \cdot {\cal B}(f_0(980) \to \pi^+ \pi^-)\; .
\eeq
Thus, one can easily obtain the $B_s^0 \to \eta_c f_0(980) (\to \pi^+ \pi^-)$ BR in the iPQCD approach as,
\beq
{\rm BR}(B_s^0 \to \eta_c f_0(980) (\to \pi^+ \pi^-)) &=& (2.87^{+1.38}_{-1.29}) \times 10^{-4}\;,
\label{eq:e2pi-nwa}
\eeq
Evidently, it is consistent with the available evidence, i.e.,
$(1.76 \pm 0.67) \times 10^{-4}$ reported by the LHCb experiment as shown
in Eq.~(\ref{eq:lhcb-e2pi}). And, theoretically, this result is also compatible
with those presented in Refs.~\cite{Ke:2017wni,Xie:2018rqv} within large
uncertainties. However, it is worth mentioning that the results obtained
through the quasi-two-body decay $B_s^0 \to \eta_c f_0(980) (\to \pi^+ \pi^-)$
in the traditional PQCD approach is about a factor of 10 smaller than the value
given in Eq.~(\ref{eq:e2pi-nwa}). The future precise measurements could
help clarify this discrepancy.

As the mixed partner of $f_0(980)$ in the two-quark picture, there are no any
available measurements of ${\cal B}(f_0(500) \to \pi^+ \pi^-)$ currently.
Then, for the sake of convenience, its value has to be taken as $0.67 \pm 0.067$
based on the assumption of $f_0(500)$ decaying totally into $\pi\pi$ associated
with isospin symmetry. Therefore, we can obtain the values of rest ${\rm BR}(B^0
\to \eta_c f_0(500, 980) (\to \pi^+ \pi^-))$ at $|\varphi| \sim 25^\circ$,
\beq
	{\rm BR}(B_d^0 \to \eta_c f_0(500) (\to \pi^+ \pi^-)) &=&
    (3.42^{+1.49}_{-1.49}) \times 10^{-5}\;,
    \\
	{\rm BR}(B_d^0 \to \eta_c f_0(980) (\to \pi^+ \pi^-)) &=&
    (0.46^{+0.23}_{-0.21}) \times 10^{-5}\;,
    \\
{\rm BR}(B_s^0 \to \eta_c f_0(500) (\to \pi^+ \pi^-)) &=&
    (1.00^{+0.50}_{-0.49}) \times 10^{-4}\;.
\eeq
One can find easily that the iPQCD result of ${\rm BR}(B_d^0 \to \eta_c f_0(500)
(\to \pi^+ \pi^-))$ is very close to that predicted in~\cite{Xie:2018rqv} within
theoretical uncertainties. Certainly, these large values around $10^{-6} \sim 10^{-4}$
would be tested at the LHCb and Belle-II experiments in the near future.

Additionally, though $f_0(980)$ coming from the $K^+ K^-$ invariant mass
is not easy to be detected currently because $f_0(980)$ is usually buried in
the tail of the meson $\phi$, we still predict the BRs of possible channels
induced by $f_0(980) \to K^+ K^-$ for future examinations
with the gradually improved techniques in detection and analysis of experimental data.
With ${\cal B}(f_0(980) \to K^+ K^-) = 0.16^{+0.04}_{-0.05}$\cite{Liu:2019ymi}, we have
\beq
{\rm BR}(B_d^0 \to \eta_c f_0(980) (\to K^+ K^-)) &=&
		(0.16^{+0.09}_{-0.09}) \times 10^{-5}\;, \\
{\rm BR}(B_s^0 \to \eta_c f_0(980) (\to K^+ K^-)) &=&
		(1.02^{+0.52}_{-0.54}) \times 10^{-4}\;.
\eeq

Then, the ratios between ${\rm BR}(B^0 \to \eta_c f_0(500) (\to \pi^+ \pi^-))$ and
${\rm BR}(B^0 \to \eta_c f_0(980) (\to \pi^+ \pi^-/K^+ K^-))$ are defined
and presented in order:
\beq
R^{\pi\pi}_{dd}[f_0(980)/f_0(500)] &\equiv&
	\frac{{\rm BR}(B_{d}^{0} \to \eta_c f_{0}(980)(\to \pi^+ \pi^-))}
	{{\rm BR}(B_{d}^{0} \to \eta_c f_{0}(500)(\to \pi^+ \pi^-))}
	\approx
	0.13^{+0.09}_{-0.08}\;,
\\
R_{ss}^{\pi\pi}[f_0(500)/f_0(980)] &\equiv&
		\frac{{\rm BR}(B_{s}^{0} \to \eta_c f_{0}(500)(\to \pi^+ \pi^-))}
		{{\rm BR}(B_{s}^{0} \to \eta_c f_{0}(980)(\to \pi^+ \pi^-))}
		\approx
		0.35^{+0.24}_{-0.23}\;,
\\
R^{\pi \pi}_{ds}[f_0(500)/f_0(980)] &\equiv&
\frac{{\rm BR}(B_{d}^{0} \to \eta_c f_{0}(500)(\to \pi^+ \pi^-))}
{{\rm BR}(B_{s}^{0} \to \eta_c f_{0}(980)(\to \pi^+ \pi^-))}
\approx
0.12^{+0.08}_{-0.07}\;,
\\
R_{ds}^{\pi\pi}[f_0(980)/f_0(500)] &\equiv&
\frac{{\rm BR}(B_{d}^{0} \to \eta_c f_{0}(980)(\to \pi^+ \pi^-))}
{{\rm BR}(B_{s}^{0} \to \eta_c f_{0}(500)(\to \pi^+ \pi^-))}
\approx
0.05^{+0.03}_{-0.03}\;,
\eeq
in which, it is clearly seen that the ratio $R^{\pi \pi}_{ds}[f_0(500)/f_0(980)]
= 0.12^{+0.08}_{-0.07}$ is consistent well with the available prediction in
Eq.~(\ref{eq:r-xie}) within uncertainties.

Moreover, we have
\beq
R_{dd}^{K\pi}
&\equiv&
	\frac{{\rm BR}(B_{d}^{0} \to \eta_c f_{0}(980)(\to K^+ K^-))}
	{{\rm BR}(B_{d}^{0} \to \eta_c f_{0}(500)(\to \pi^+ \pi^-))}
	\approx
	0.05^{+0.03}_{-0.03}\,
\\
R^{K \pi}_{ds}
&\equiv&
\frac{{\rm BR}(B_{d}^{0} \to \eta_c f_{0} (980)(\to K^+ K^-))}
{{\rm BR}(B_{s}^{0} \to \eta_c f_{0} (500)(\to \pi^+ \pi^-))}
\approx
0.02^{+0.01}_{-0.01}\; ,
\\
R_{ss}^{K \pi}
&\equiv&
		\frac{{\rm BR}(B_{s}^{0} \to \eta_c f_{0}(980)(\to K^+ K^-))}
		{{\rm BR}(B_{s}^{0} \to \eta_c f_{0}(500)(\to \pi^+ \pi^-))}
		\approx
		1.02^{+0.73}_{-0.74}\;,
\\
R_{sd}^{K \pi} &\equiv&
\frac{{\rm BR}(B_{s}^{0} \to \eta_c f_{0} (980)(\to K^+ K^-))}
{{\rm BR}(B_{d}^{0} \to \eta_c f_{0} (500)(\to \pi^+ \pi^-))}
\approx
2.98^{+2.00}_{-2.04} \;.
\eeq
All the above ratios await experimental tests in the future
to further decipher the QCD dynamics of $f_0(500, 980)$.

\begin{table}[htb]
	\caption{Same as Table~\ref{tab:BR-ls} but for $B^0 \to \eta_c f_0(1370,1500)$, where the
1st (2nd) entry in every line corresponds to $f_0$ in $S1 (S2)$. }
	\label{tab:BR-hs}
	\begin{center}\vspace{-0.5cm}{
			\begin{tabular}[t]{c| c}
				\hline \hline
				Decay modes   &  Branching ratios  \\
				\hline
				$B_d^0 \to \eta_c f_0(1370)$
				&
$\begin{array}{cc}
4.08^{+1.83}_{-1.21}(\omega_{B^0})
				^{+0.88}_{-1.19}(f_{\eta_c})
                ^{+0.92}_{-0.82}(\bar{f}_{f'_{0n}})
				^{+5.43}_{-2.98}(B_{i}^{n'})
				^{+0.00}_{-0.42}(a_t)
				^{+0.71}_{-0.80}(\varphi')
				\times 10^{-6}
\\
4.57^{+1.45}_{-1.05}(\omega_{B^0})
				^{+0.98}_{-1.33}(f_{\eta_c})
                ^{+1.04}_{-0.94}(\bar{f}_{f'_{0n}})
				^{+0.94}_{-0.88}(B_{i}^{n'})
				^{+0.20}_{-0.36}(a_t)
				^{+0.80}_{-0.90}(\varphi')
				\times 10^{-5}
\end{array}$		
\\			
\hline		
	$B_d^0 \to \eta_c f_0(1500)$
				&
$\begin{array}{cc}
1.72^{+0.78}_{-0.52}(\omega_{B^0})
				^{+0.37}_{-0.50}(f_{\eta_c})
                ^{+0.39}_{-0.35}(\bar{f}_{f'_{0n}})
				^{+2.30}_{-1.26}(B_{i}^{n'})
				^{+0.00}_{-0.18}(a_t)
				^{+0.75}_{-0.66}(\varphi')
				\times 10^{-6}
\\
1.94^{+0.61}_{-0.45}(\omega_{B^0})
				^{+0.41}_{-0.57}(f_{\eta_c})
                ^{+0.44}_{-0.40}(\bar{f}_{f'_{0n}})
				^{+0.40}_{-0.38}(B_{i}^{n'})
				^{+0.08}_{-0.15}(a_t)
				^{+0.83}_{-0.75}(\varphi')
				\times 10^{-5}
\end{array} $
				\\
\hline  \hline
$B_s^0 \to \eta_c f_0(1370)$
				&
$\begin{array}{cc}
1.91^{+0.99}_{-0.63}(\omega_{B^0})
				^{+0.41}_{-0.56}(f_{\eta_c})
                ^{+0.43}_{-0.39}(\bar{f}_{f'_{0s}})
				^{+3.23}_{-1.53}(B_{i}^{s'})
				^{+0.08}_{-0.22}(a_t)
				^{+0.82}_{-0.74}(\varphi')
				\times 10^{-5}
\\
2.68^{+0.95}_{-0.67}(\omega_{B^0})
				^{+0.58}_{-0.79}(f_{\eta_c})
                ^{+0.61}_{-0.55}(\bar{f}_{f'_{0s}})
				^{+0.60}_{-0.56}(B_{i}^{s'})
				^{+0.15}_{-0.21}(a_t)
				^{+1.15}_{-1.04}(\varphi')
				\times 10^{-4}
\end{array}$
				\\   \hline
$B_s^0 \to \eta_c f_0(1500)$
				&
$\begin{array}{cc}
3.93^{+2.05}_{-1.29}(\omega_{B^0})
				^{+0.85}_{-1.15}(f_{\eta_c})
                ^{+0.89}_{-0.80}(\bar{f}_{f'_{0s}})
				^{+6.65}_{-3.15}(B_{i}^{s'})
				^{+0.16}_{-0.44}(a_t)
				^{+0.69}_{-0.77}(\varphi')
				\times 10^{-5}
\\
5.53^{+1.97}_{-1.39}(\omega_{B^0})
				^{+1.19}_{-1.61}(f_{\eta_c})
                ^{+1.27}_{-1.14}(\bar{f}_{f'_{0s}})
				^{+1.25}_{-1.15}(B_{i}^{s'})
				^{+0.32}_{-0.43}(a_t)
				^{+0.97}_{-1.09}(\varphi')
				\times 10^{-4}
\end{array}$
		\\		
\hline \hline
		\end{tabular}}
	\end{center}
\end{table}	

Let's turn to the decays of $B^0 \to \eta_c f_0(1370, 1500)$. Again,
$f_0(1370, 1500)$ are thought as the pure scalar mesons in two different scenarios.
Therefore, the 1st (2nd) entry of numerical results displayed in every line of
Table~\ref{tab:BR-hs} corresponds to $f_0(1370, 1500)$ in $S1 (S2)$.
Table~\ref{tab:BR-hs} contains the iPQCD predictions of
$B^{0} \to \eta_c f_0(1370, 1500)$ BRs at $ \vert \varphi^\prime \vert
= (146 \pm 8)^\circ $ with uncertainties from various sources. Obviously,
the BRs predicted in $S1$ are globally smaller than those in $S2$ about an order.
Frankly speaking, the largest uncertainties result from the least constrained hadronic
parameters, i.e., Gegenbauer moments $B_i^{n^\prime, s^\prime}$,
as well as scalar decay constants $\bar f_{f'_{0n(s)}}$ in the LCDAs, which are
nonperturbative but key inputs in the QCD-based factorization approach.
Nevertheless, such large BRs around $10^{-6} \sim 10^{-4}$ could be accessed
promisingly in the near-future LHCb and Belle-II experiments. Experimental tests
of these iPQCD predictions can hint at the magnitude of mixing angle $\varphi^\prime$
between $f_{0}(1370)$ and $f_0(1500)$. In principle, it could help us to identify the
scalar glueball components involved in these two mesons, which would potentially
provide evidences to help differentiate $f_0(1500)$ as a primary or
fragmented scalar glueball.

Analogous to the decays $B^0 \to \eta_c f_0(500, 980)$, more ratios between
the $B^0 \to \eta_c f_0(1370, 1500)$ BRs could be derived to provide helpful
hints for the involved mixing angle $\varphi^\prime$ and/or broken SU(3)
flavor symmetries. Certainly, future tests from various experiments would
even tell us the preferred scenario of $f_0(1370, 1500)$ that could help us
to further understand the QCD dynamics or internal structure of light scalars.
\beq
R_{d}[f_0(1370)/f_0(1500)] &\equiv&
	\frac{{\rm BR}(B_{d}^{0} \to \eta_c f_{0}(1370))}
	{{\rm BR}(B_{d}^{0} \to \eta_c f_{0}(1500))}
	= \frac{\Phi^{d}_{f_0(1370)}}{\Phi^{d}_{f_0(1500)}}
\cdot \cot^2 \varphi^\prime
	\approx
	\biggl\{\begin{array}{rr}
    2.37^{+0.72}_{-0.43}
            \; (S1)
		\\
		2.36^{+0.72}_{-0.42}
        \; (S2)
	\end{array}\;,
\\	
R_{s}[f_0(1500)/f_0(1370)] &\equiv&
\frac{{\rm BR}(B_{s}^{0} \to \eta_c f_{0}(1500)}
{{\rm BR}(B_{s}^{0} \to \eta_c f_{0}(1370))}
= \frac{\Phi^{s}_{f_0(1500)}}{\Phi^{s}_{f_0(1370)}}
\cdot \cot^2 \varphi^\prime
\approx
	\biggl\{\begin{array}{rr}
	2.06^{+0.64}_{-0.37}
    \; (S1)
	\\
   2.06^{+0.65}_{-0.36}
    \; (S2)
\end{array}\;,
\eeq
\beq
R_{d}^{s}[f_0(1370)] &\equiv&
\frac{{\rm BR}(B_{s}^{0} \to \eta_c f_{0}(1370))}
{{\rm BR}(B_{d}^{0} \to \eta_c f_{0}(1370))}
\non
&=& \frac{\tau_{B_s^0}}{\tau_{B_d^0}}\cdot (\frac{m_{B_s^0}}{m_{B_d^0}})^7
\cdot \frac{\Phi^{s}_{f_0(1370)}}{\Phi^{d}_{f_0(1370)}}
\cdot \frac{|{\cal A}(B_s^0 \to \eta_c f'_{0s})/m^2_{B_s^0}|^2}
{|{\cal A}(B_d^0 \to \eta_c f'_{0n})/m^2_{B_d^0}|^2}
\cdot \tan^2 \varphi^\prime
\approx
	\biggl\{\begin{array}{rr}
    4.68^{+1.30}_{-1.53}
    \; (S1)
	\\
   5.86^{+1.29}_{-1.40}
    \; (S2)
\end{array}\;,
\eeq
\beq
R_{d}^{s}[f_0(1500)] &\equiv&
\frac{{\rm BR}(B_{s}^{0} \to \eta_c f_{0}(1500))}
{{\rm BR}(B_{d}^{0} \to \eta_c f_{0}(1500))}
\non
&=& \frac{\tau_{B_s^0}}{\tau_{B_d^0}}\cdot (\frac{m_{B_s^0}}{m_{B_d^0}})^7
\cdot \frac{\Phi^{s}_{f_0(1500)}}{\Phi^{d}_{f_0(1500)}}
\cdot \frac{|{\cal A}(B_s^0 \to \eta_c f'_{0s})/m^2_{B_s^0}|^2}
{|{\cal A}(B_d^0 \to \eta_c f'_{0n})/m^2_{B_d^0}|^2}
\cdot \cot^2 \varphi^\prime
\approx
	\biggl\{\begin{array}{rr}
	22.85^{+7.94}_{-6.65}
    \; (S1)
	\\
   28.51^{+8.87}_{-5.12}
    \; (S2)
\end{array}\;,
\eeq
\beq
R_{d}^{s}[f_0(1500)/f_0(1370)] &\equiv&
\frac{{\rm BR}(B_{s}^{0} \to \eta_c f_{0}(1500))}
{{\rm BR}(B_{d}^{0} \to \eta_c f_{0}(1370))}
\non
&=& \frac{\tau_{B_s^0}}{\tau_{B_d^0}}\cdot (\frac{m_{B_s^0}}{m_{B_d^0}})^7
\cdot \frac{\Phi^{s}_{f_0(1500)}}{\Phi^{d}_{f_0(1370)}}
\cdot \frac{|{\cal A}(B_s^0 \to \eta_c f'_{0s})/m^2_{B_s^0}|^2}
{|{\cal A}(B_d^0 \to \eta_c f'_{0n})/m^2_{B_d^0}|^2}
\approx
\biggl\{\begin{array}{rr}
9.63^{+1.66}_{-2.14}
\; (S1)
	\\
   12.10^{+0.44}_{-0.42}
\; (S2)
\end{array}\;,
\\
R_{d}^{s}[f_0(1370)/f_0(1500)] &\equiv&
\frac{{\rm BR}(B_{s}^{0} \to \eta_c f_{0}(1370))}
{{\rm BR}(B_{d}^{0} \to \eta_c f_{0}(1500))}
\non
&=& \frac{\tau_{B_s^0}}{\tau_{B_d^0}}\cdot (\frac{m_{B_s^0}}{m_{B_d^0}})^7
\cdot \frac{\Phi^{s}_{f_0(1370)}}{\Phi^{d}_{f_0(1500)}}
\cdot \frac{|{\cal A}(B_s^0 \to \eta_c f'_{0s})/m^2_{B_s^0}|^2}
{|{\cal A}(B_d^0 \to \eta_c f'_{0n})/m^2_{B_d^0}|^2}
\approx
\biggl\{\begin{array}{rr}
11.10^{+1.85}_{-2.57}
\; (S1)
	\\
 13.81^{+0.52}_{-0.40}
    \; (S2)
\end{array}\;,
\eeq
where all the uncertainties from various sources have been added in quadrature.

Two comments are in order:
(a)
The first two ratios are scenario independent naturally because same transitions
$B_{d(s)}^0 \to f^\prime_{0n(s)}$ are contained simultaneously in the numerator and denominator,
which leads to an exact cancelation of the uncertainties induced by the same
hadronic parameters. Thus, they are determined by only the mixing angle $\varphi^\prime$.
It means that, once these two ratios are measured precisely, then the magnitude of
$\varphi^\prime$ could be determined cleanly with the help of definite values
$\Phi_{d}^{f_0(1370)} /{\Phi_{d}^{f_0(1500)}} \approx 1.075$, and $\Phi_{s}^{f_0(1370)}
/ {\Phi_{s}^{f_0(1500)}} \approx 1.067$, and vice versa.
(b)
Similar patterns occur in the last two ratios, that is, both values in $S1$ are close
to while slightly less than those in $S2$. The future stringent tests with precise
data might indicate the preferred one of those two scenarios classifying the light
scalars $f_0(1370, 1500)$. It means that, inferred from the last two ratios describing
the effects of broken SU(3) flavor symmetries, we could get some evidences to further
study their QCD dynamics.

It is known that the LHCb experiment has observed $B_s^0 \to J/\psi f_0(1500)$ through
measuring the BR with secondary decay chain $B_{s}^0 \to J/\psi f_0(1500) ( \to \pi^+ \pi^-)$
as $(2.04^{+0.32}_{-0.24}) \times 10^{-5}$~\cite{Workman:2022pdg}. It is then expected
that the upgraded LHCb detector could help observe the decay $B_s^0 \to \eta_c f_0(1500)
(\to \pi^+ \pi^-)$ promisingly in the near future. For convenient and effective comparisons
with future measurements of the decays $B^0 \to \eta_c f_0(1500)$, we derive the three-body
$B^0 \to \eta_c f_0(1500) (\to \pi^+ \pi^-)$ BRs in the iPQCD formalism by employing the data
${\cal B}(f_0(1500) \to \pi \pi)=(34.5\pm2.2)\% $~\cite{Workman:2022pdg} under NWA~\cite{Cheng:2015iaa}.
With ${\cal B}(f_0(1500) \to \pi^+ \pi^-)=0.230\pm 0.015$ and $|\varphi^\prime|=146^\circ$, we have
\beq
{\rm BR}(B_d^0 \to \eta_c f_0(1500) (\to \pi^+ \pi^-)) =
\biggl\{\begin{array}{rr}
0.40^{+0.59}_{-0.38} \times 10^{-6}
\; (S1)
\\
0.45^{+0.29}_{-0.28} \times 10^{-5}
\; (S2)
\end{array}\;,		
\\
{\rm BR}(B_s^0 \to \eta_c f_0(1500) (\to \pi^+ \pi^-)) =
	\biggl\{\begin{array}{rr}
    0.90^{+1.64}_{-0.89}
		\times 10^{-5}
\; (S1)
		\\
     1.27^{+0.71}_{-0.67}
	\times 10^{-4}
\; (S2)
	\end{array}\;,
\eeq
It is noted that, within the remarkably large
errors, our numerical result of ${\rm BR}(B_s^0 \to \eta_c f_0(1500)(\to \pi^+ \pi^-))$
in $S1$ is consistent with the value obtained in the traditional PQCD approach through
quasi-two-body investigations~\cite{Li:2015tja}. These predictions are expected to be
confronted with the near future LHCb and Belle-II experiments, which will provide
hints to identify the preferred scenario, which will help to study its QCD dynamics
and further distinguish $f_0(1500)$ as a primary or fragmented scalar glueball.

Meanwhile, according to the data ${\cal B}(f_0(1500) \to K^{+}K^{-})=(4.25\pm 0.50)\%$~\cite{Workman:2022pdg},
we can also derive the $B_{s}^0 \to \eta_c f_0(1500) (\to K^{+} K^{-})$ BRs in the iPQCD
formalism under NWA at $|\varphi^\prime| = 146^\circ$,
\beq
	{\rm BR}(B_d^0 \to \eta_c f_0(1500) (\to K^+ K^-)) &=&
\biggl\{\begin{array}{rr}
0.07^{+0.10}_{-0.07}\times 10^{-6}
\; (S1)
\\
0.08^{+0.05}_{-0.05} \times 10^{-5}
\; (S2)
\end{array}\;,	
\\
{\rm BR}(B_s^0 \to \eta_c f_0(1500) (\to K^+ K^-)) &=&
\biggl\{\begin{array}{rr}
0.16^{+0.28}_{-0.16}\times 10^{-5}
\; (S1)
\\
0.22^{+0.13}_{-0.13}\times 10^{-4}
\; (S2)
\end{array}\;, 	
\eeq
The $B_s^0 \to \eta_c f_0(1500)(\to K^+ K^-)$ BRs 
are expected to be accessible in the LHCb and Belle-II experiments,
which can also help us to understand the QCD dynamics of $f_0(1500)$.

\begin{figure}[!!htb]
\begin{center}
\includegraphics[scale=0.6]{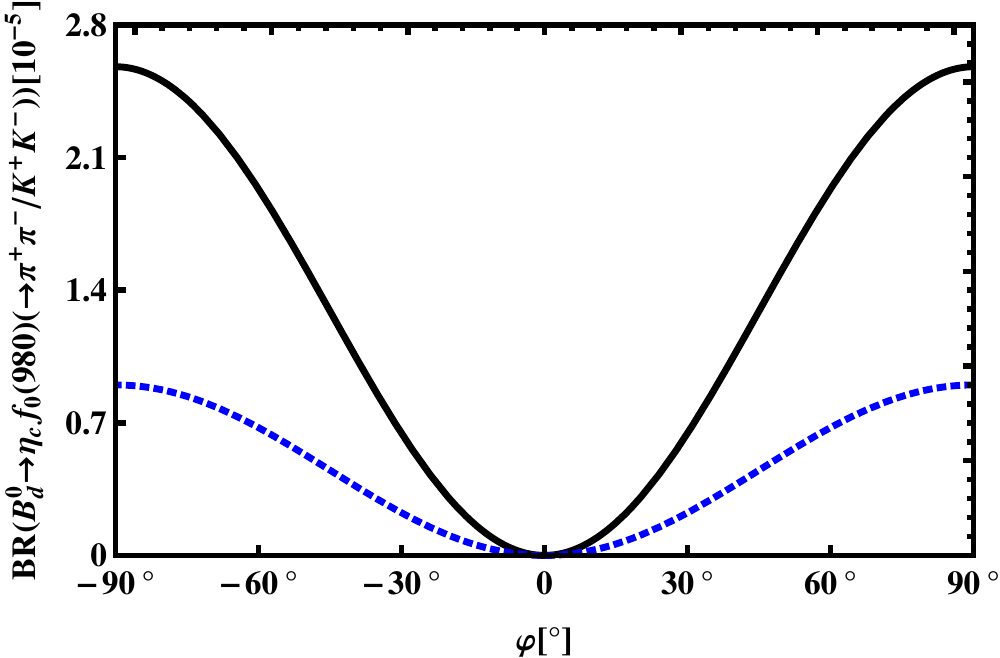}\hspace{0.8cm}
\includegraphics[scale=0.6]{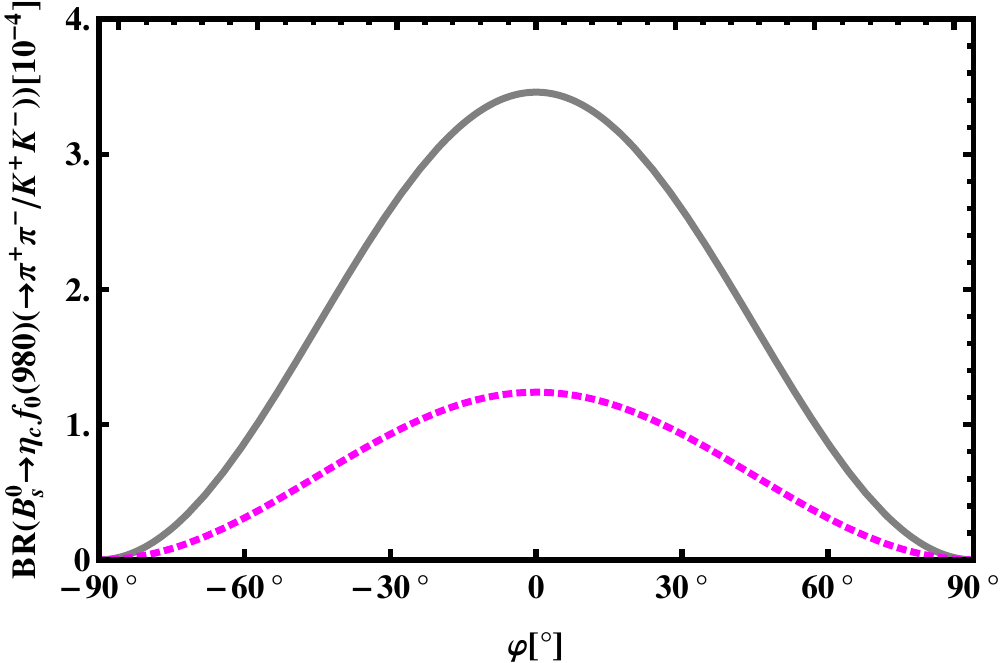}
		\caption{(Color online)  Dependence
of ${\rm BR}(B^{0} \to \eta_c f_{0}(980) (\to \pi^+\pi^-/K^+K^-))$ on $\varphi \in [-90^\circ, 90^\circ]$ in the iPQCD formalism:
solid (dotted) line corresponds to $B^{0}
\to \eta_c f_{0}(980) (\to \pi^+\pi^-) [B^0 \to \eta_c f_0(980) (\to K^+ K^-)]$.}
		\label{fig:fig4}
	\end{center}
\end{figure}

\begin{figure}[!!htb]
\begin{center}
\includegraphics[scale=0.6]{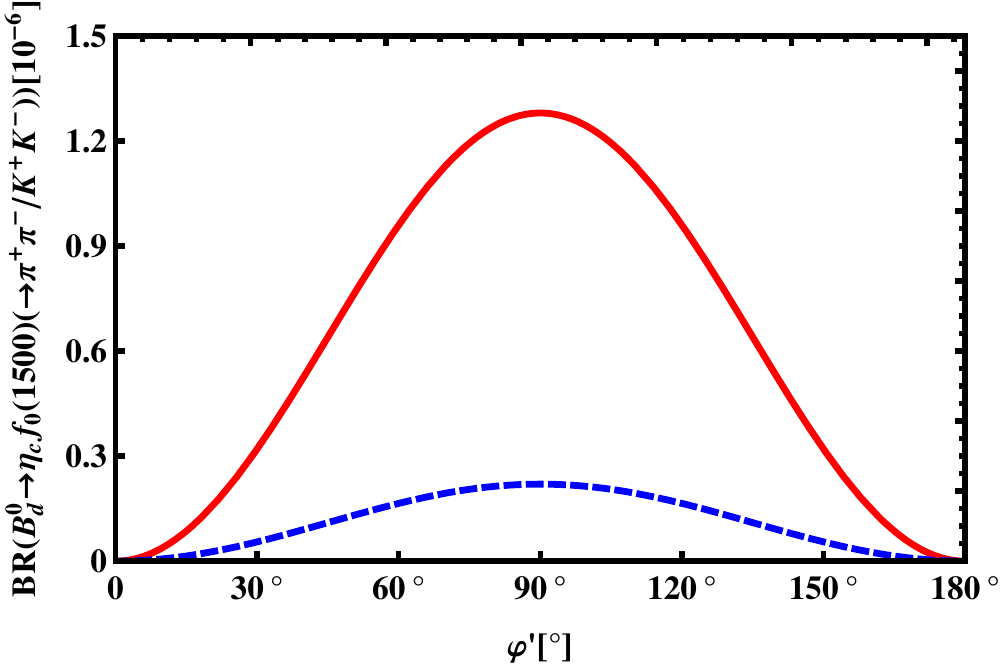}\hspace{0.8cm}
\includegraphics[scale=0.6]{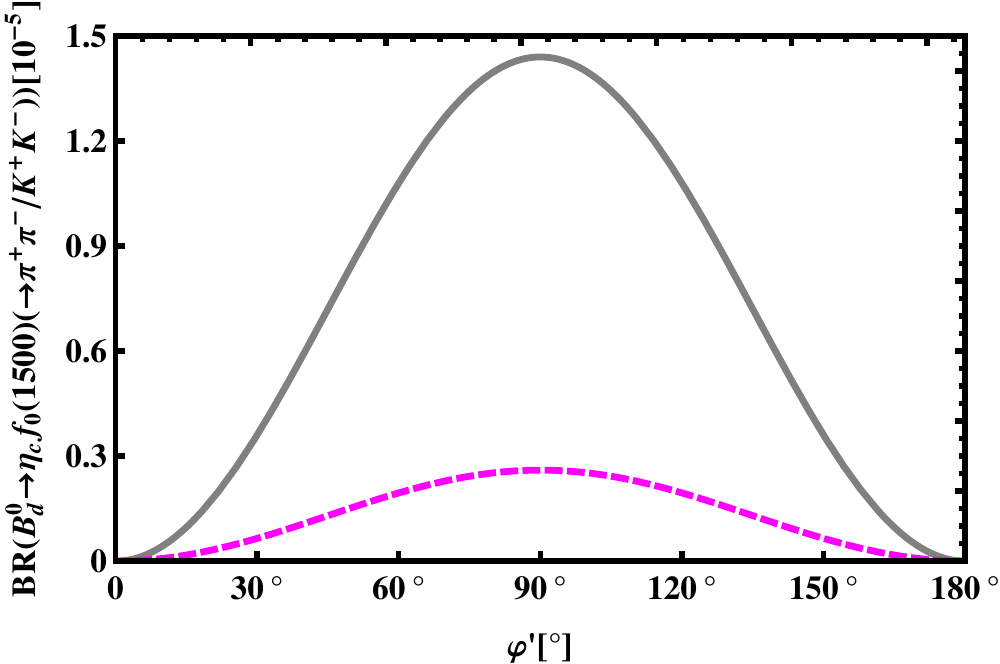}
		\caption{(Color online)  Dependence
of ${\rm BR}(B_d^{0} \to \eta_c f_{0}(1500) (\to \pi^+\pi^-/K^+K^-))$ on $\varphi^\prime \in [0^\circ, 180^\circ]$ in the iPQCD formalism:
solid (dashed) line corresponds to $B_{d}^{0}
\to \eta_c f_{0}(1500) (\to \pi^+ \pi^-) [B_{d}^{0} \to \eta_c f_{0}(1500) (\to K^+K^-)]$ with
left (right) panel in $S1 (S2)$.}
		\label{fig:fig5}
	\end{center}
\end{figure}

\begin{figure}[!!htb]
\begin{center}
\includegraphics[scale=0.6]{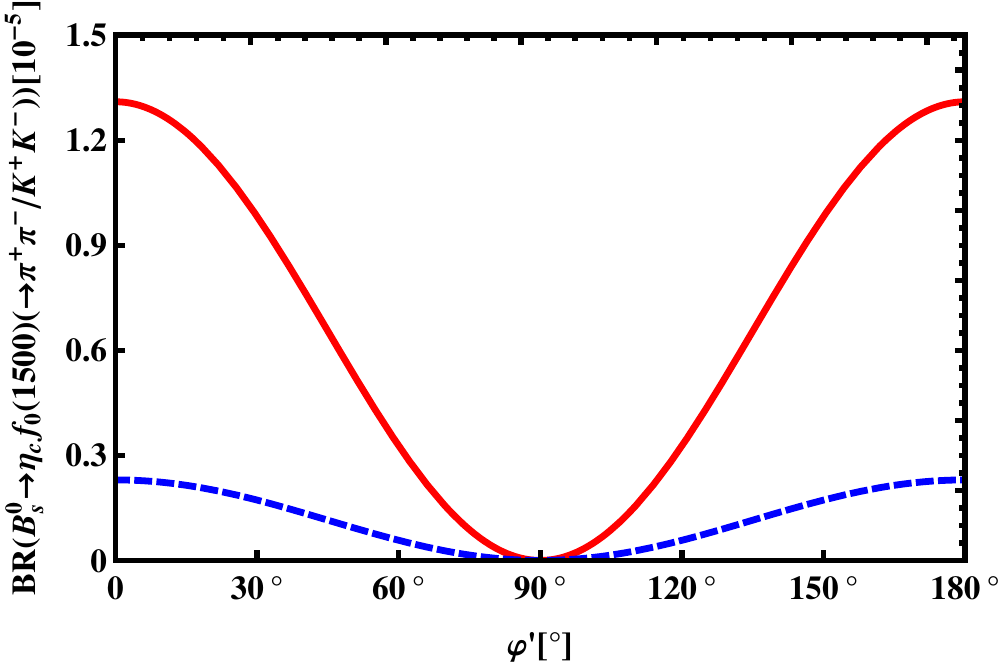}\hspace{0.8cm}
\includegraphics[scale=0.6]{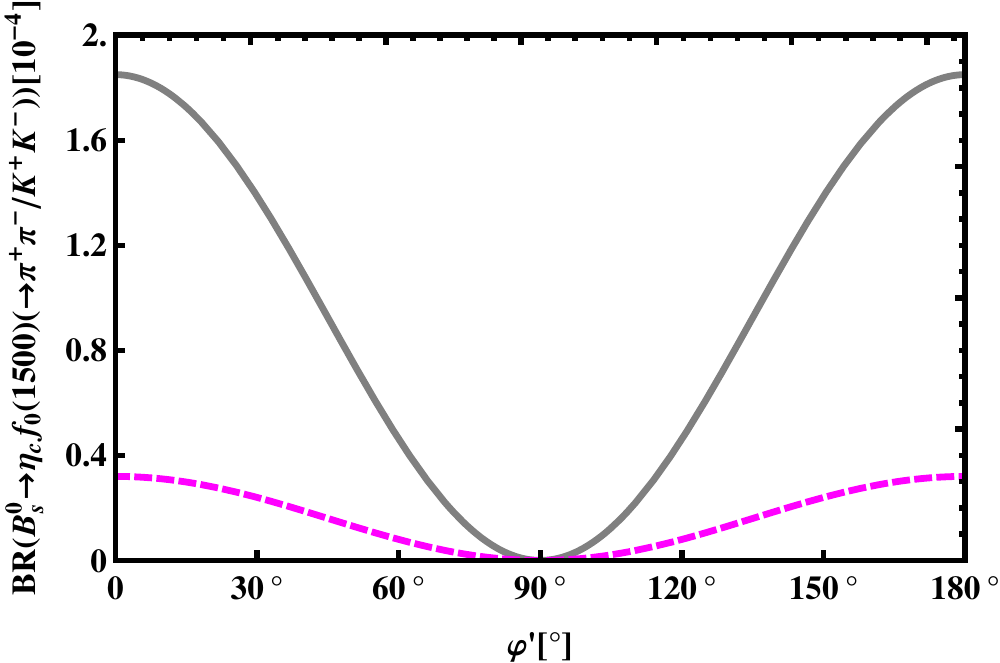}
		\caption{(Color online)  Dependence
of ${\rm BR}(B_s^{0} \to \eta_c f_{0}(1500) (\to \pi^+\pi^-/K^+K^-))$ on $\varphi^\prime \in [0^\circ, 180^\circ]$ in the iPQCD formalism:
solid (dashed) line corresponds to $B_{s}^{0}
\to \eta_c f_{0}(1500) (\to \pi^+ \pi^-) [B_{s}^{0} \to \eta_c f_{0}(1500) (\to K^+K^-)]$ with
left (right) panel in $S1 (S2)$.}
		\label{fig:fig6}
	\end{center}
\end{figure}

To provide more information of $\varphi^{(\prime)}$ for reference,
the BRs of $B^0 \to \eta_c f_0(980, 1500) (\to \pi^+ \pi^-/ K^+ K^-)$ varying
with mixing angle are plotted and presented in Figs.~\ref{fig:fig4}-\ref{fig:fig6}.
The experimental tests of ${\rm BR}(B_s^0 \to \eta_c f_0(980, 1500) (\to \pi^+\pi^-
/ K^+ K^-))$ would be helpful to further explore the $s\bar s$ components, as well
as to constrain their mixing angle.

Finally, we turn to analyze CPAs of the $B^{0} \to \eta_c f_0$ decays in the iPQCD
approach at NLO accuracy. For the CPAs in these considered $B^{0}$ decays, the $B^{0}
- \bar B^{0}$ mixing effects should be taken into account. As is well known, the CPA
of the $B^{0}(\bar B^{0}) \to \eta_c f_0$ decays is time-dependent and can be defined
as
\beq
A_{\rm CP} &\equiv& \frac{\Gamma\left
	(\bar{B}^0(\Delta t) \to f_{\rm CP}\right) -
	\Gamma\left(B^0(\Delta t) \to f_{\rm CP}\right )}{ \Gamma\left
	(\bar{B}^0(\Delta t) \to f_{\rm CP}\right ) + \Gamma\left
	(B^0(\Delta t) \to f_{\rm CP}\right ) }\non
&=& A_{\rm CP}^{\rm dir} \cos(\Delta m  \Delta t)
+ A_{\rm CP}^{\rm mix} \sin (\Delta m \Delta
t)\;, \label{eq:acp-def}
\eeq
where $\Delta m$ is the mass difference between the two $B^0$ mass eigenstates,
$\Delta t = t_{\rm CP}-t_{tag} $ is the time difference between the tagged
$B^0$ ($\bar{B}^0$) and the accompanying $\bar{B}^0$ ($B^0$) with opposite
$b$ flavor decaying to the final CP-eigenstate $f_{\rm CP}$ at the time $t_{\rm CP}$.
The direct and mixing-induced CPAs $A_{\rm CP}^{\rm dir}$ and $A_{\rm CP}^{\rm mix}$
can be written as
\beq
A_{\rm CP}^{\rm dir}&\equiv&
\frac{ \left |
	\lambda_{\rm CP}^{d,s}\right |^2-1 } {1+\left |\lambda_{\rm CP}^{d,s}\right |^2},
\qquad
A_{\rm CP}^{\rm mix}\equiv
\frac{ 2 {\rm Im}
	(\lambda_{\rm CP}^{d,s})}{1+\left |\lambda_{\rm CP}^{d,s}\right |^2},
\label{eq:acp-csf}
\eeq
where the {\it CP}-violating parameters $\lambda_{\rm CP}^{d,s}$ can be read as
\beq
\lambda_{\rm CP}^{d} &\equiv& \eta_f \; \frac{V_{tb}^*V_{td}}{V_{tb}V_{td}^*}
\cdot \frac{ \langle f_{\rm CP} |H_{\rm eff}|\bar{B}_{d}^0\rangle}
{\langle f_{\rm CP} |H_{\rm eff}|B_{d}^0\rangle},
\qquad
\lambda_{\rm CP}^{s} \equiv \eta_f \; \frac{V_{tb}^*V_{ts}}{V_{tb}V_{ts}^*}
\cdot \frac{ \langle f_{\rm CP} |H_{\rm eff}|\bar{B}_{s}^0\rangle}
{\langle f_{\rm CP} |H_{\rm eff}|B_{s}^0\rangle},
\label{eq:lambda-ds}
\eeq
with {\it CP}-eigenvalue of the final states $\eta_f = +1$. Notice that,
for the $B_s^0$-meson decays, due to the presence of a non-negligible
$\Delta \Gamma_s$, a non-zero ratio $(\Delta \Gamma/\Gamma)_{B_s^0}$
is expected in the standard model~\cite{Beneke:1998sy,Fernandez:2006qx}.
Thus, for $B_s^0 \to \eta_c f_0$ decays, the $\Delta \Gamma_s$-induced
CPA $A_{\rm CP}^{\Delta \Gamma_s}$ can be defined as follows~\cite{Fernandez:2006qx}:
\beq
A_{\rm CP}^{\Delta \Gamma_s} &\equiv& \frac{ 2 {\rm Re}
	( \lambda_{\rm CP}^{s})}{1+\left |\lambda_{\rm CP}^{s}\right |^2}.
\label{eq:acp-dgs}
\eeq
The above three CPAs in $B_s^0$-meson decays shown in Eqs.~(\ref{eq:acp-csf})
and (\ref{eq:acp-dgs}) satisfy the following relation,
\beq
|A_{\rm CP}^{\rm dir}|^2+ |A_{\rm CP}^{\rm mix}|^2
+ |A_{\rm CP}^{\Delta \Gamma_s}|^2 &=&
1 \;.\label{eq:summation-cp}
\eeq

\begin{table}[htb]
	\caption{ CPAs of $B^0 \to \eta_c f_0(500, 980)$ in iPQCD formalism. }
	\label{tab:CPV-f980}
	\begin{center}\vspace{-0.5cm}{
			\begin{tabular}[t]{c| c |c |c}
				\hline \hline
				Decay modes   & $A_{\rm CP}^{\rm dir}$  &  $A_{\rm CP}^{\rm mix}$
& $A_{\rm CP}^{\rm \Delta \Gamma_s}$
   \\
				\hline
				$B_d^0 \to \eta_c f_{0n}$ &
              $(-2.59^{+0.63}_{-0.39})\times 10^{-2}$ &
              $(-69.20^{+1.77}_{-1.72}) \times 10^{-2}$ &
              $-$
\\
\hline
			    $B_s^0 \to \eta_c f_{0s}$ &
              $(1.28^{+0.22}_{-0.35}) \times 10^{-3}$ &
              $(3.50^{+0.16}_{-0.12}) \times 10^{-2}$ &
              $0.999$
				\\
				\hline \hline
		\end{tabular}}
	\end{center}
\end{table}	

The CPAs $A_{\rm CP}^{\rm dir}$, $A_{\rm CP}^{\rm mix}$ and $A_{\rm CP}^{\rm \Delta
\Gamma_s}$ of the $B^0 \to \eta_c f_0$ decays in the iPQCD formalism are collected
in Tables~\ref{tab:CPV-f980} and~\ref{tab:CPV-f1315}. All the uncertainties from
various sources have been added in quadrature. Since the $B_d^0 \to \eta_c f_0$
and $B_s^0 \to \eta_c f_0$ decays are governed by the $B_d^0 \to f_{0n}^{(\prime)}$
and $B_s^0 \to f_{0s}^{(\prime)}$ transitions, respectively, then their CPAs are
determined by the dynamics of $B_d^0 \to \eta_c f_{0n}^{(\prime)}$ and $B_s^0 \to
\eta_c f_{0s}^{(\prime)}$ correspondingly.

Generally speaking, one can observe that, due to the very small $A_{\rm CP}^{\rm dir}$
except for that of $B_{d(s)}^0 \to \eta_c f_{0n(s)}^\prime$ in $S1$, numerical values of
$A_{\rm CP}^{\rm mix}$ in the rest decays are approximately proportional to
$\sin2\beta$ or $\sin2\beta_s$ within errors, which means few contaminations
arising from the negligible penguin contributions in the related decays.
However, as seen in Table~\ref{tab:CPV-f1315}, the remarkably large $A_{\rm CP}^{\rm dir}$
from 1st entry in every line imply the considerable penguin amplitudes contributed from
$B^0 \to \eta_c f_0(1370, 1500)$ with $f_0$ in $S1$, though these numerical results
suffer from still large uncertainties induced by not well constrained inputs. Unfortunately,
there are no any available measurements about these CPAs currently. Therefore, these iPQCD
predictions of CPAs await the experimental tests in the near future, which may deepen our
understanding of the related QCD dynamics contained in light scalars $f_0$ and the associated
decay channels.

\begin{table}[htb]
	\caption{ Same as Table~\ref{tab:CPV-f980} but for $B^0 \to \eta_c f_0(1370, 1500)$, where
the 1st (2nd) entry in every line corresponds to
$f_0$ in $S1(S2)$. }
	\label{tab:CPV-f1315}
	\begin{center}\vspace{-0.5cm}{
			\begin{tabular}[t]{c| c |c |c}
				\hline \hline
				Decay modes   & $A_{\rm CP}^{\rm dir}$  &  $A_{\rm CP}^{\rm mix}$
& $A_{\rm CP}^{\rm \Delta \Gamma_s}$
   \\
				\hline
				$B_d^0 \to \eta_c f'_{0n}$ &
              $\begin{array}{cc}
              (23.88^{+33.91}_{-13.68})\times 10^{-2}
              \\
              (-1.04^{+1.07}_{-0.94})\times 10^{-2}
              \end{array}$ &
              $\begin{array}{cc}
              (-73.19^{+19.19}_{-1.48})\times 10^{-2}
              \\
              (-71.31^{+1.82}_{-1.79})\times 10^{-2}
              \end{array}$ &
              $-$
\\
\hline
			    $B_s^0 \to \eta_c f'_{0s}$ &
              $\begin{array}{cc}
              (-1.49^{+0.91}_{-2.43})\times 10^{-2}
              \\
               (5.58^{+4.91}_{-5.93})\times 10^{-4}
              \end{array}$ &
              $\begin{array}{cc}
              (3.97^{+0.15}_{-1.91})\times 10^{-2}
              \\
              (3.66^{+0.13}_{-0.13})\times 10^{-2}
              \end{array}$ &
              $\begin{array}{cc}
              0.999
              \\
              0.999
              \end{array}$
				\\
				\hline \hline
		\end{tabular}}
	\end{center}
\end{table}	

%
\section{Conclusions and summary}
\label{sect:C&S}
We have studied the $B^0 \to \eta_c f_0$ decays in the iPQCD approach by including
the currently known NLO corrections. Here, $f_0$ denotes the scalar mesons $f_0(500,
980, 1370, 1500)$ under the assumptions of two-quark structure, in which $f_0(500)
(f_0(1370))$ and $f_0(980) (f_0(1500))$ have a mixing with angle $\varphi (\varphi^\prime)$
in the quark-flavor basis. We calculated the {\it CP}-averaged BRs and CPAs of
$B^0 \to \eta_c f_0$ and also derived the $B^0 \to \eta_c f_0(\to \pi^+\pi^-/K^+K^-)$
BRs under the narrow-width approximation at $|\varphi|=25^\circ$ and $|\varphi^\prime|
=146^\circ$ for experimental tests in the future. In summary, we found that:
\begin{itemize}
\item[(1)]
the numerical results of ${\rm BR}(B_s^0 \to \eta_c f_0(980)(\to \pi^+\pi^-)) = (2.87^{+1.38}_{-1.29})
\times 10^{-4}$ and ${\rm BR}(B_d^0 \to \eta_c f_0(500)(\to \pi^+\pi^-))/{\rm BR}(B_s^0 \to \eta_c
f_0(980)(\to \pi^+\pi^-)) = (12^{+8}_{-7})\%$ in the iPQCD formalism are consistent with the available
predictions and/or measurements within uncertainties.

\item[(2)]
the large $B_s^0 \to \eta_c f_0(1500) (\to \pi^+ \pi^-/K^+ K^-)$ BRs in the order of $10^{-6} \sim 10^{-4}$
are accessible in the near-future LHCb and Belle-II experiments. The relevant tests and verifications could
provide useful information for identifying $f_0(1500)$ as a primary or fragmented scalar glueball potentially.

\item[(3)]
the large $A_{\rm CP}^{\rm dir}(B_d^0 \to \eta_c f_0(1370, 1500)) = (23.88^{+33.91}_{-13.68})\times 10^{-2}$
and $A_{\rm CP}^{\rm dir}(B_s^0 \to \eta_c f_0(1370, 1500)) = (-1.49^{+0.91}_{-2.43})\times 10^{-2}$ imply
the important penguin contributions from the $B^0 \to \eta_c f_0(1370, 1500)$ decays in $S1$, which will
be confronted with future experiments to further understand the QCD dynamics of $f_0(1370, 1500)$.

\item[(4)]
the ratios $R_{d,s}[f_0(980)/f_0(500)]$ and $R_{d,s}[f_0(1500)/f_0(1370)]$ might provide the best ways
to explore and constrain the magnitudes of $\varphi$ and $\varphi^\prime$ in a clean manner correspondingly.

\end{itemize}


\begin{acknowledgments}
X.L. thanks Professors Hai-Yang Cheng and Qiang Zhao for valuable discussions on the scalar meson
and scalar glueball. M.L. thanks D.Y. and J.R. for their helpful discussions. This work is supported
by the National Natural Science Foundation of China under the Grants Nos.~11875033,~12335003,~12375089, and~12435004,
by the Qing Lan Project of Jiangsu Province (9212218405), and by the Natural Science Foundation
of Shandong province under the Grants Nos.~ZR2022ZD26 and~ZR2022MA035.
M.L. is supported by Postgraduate Research $\&$ Practice Innovation Program of Jiangsu Normal
University (2022XKT1327).
\end{acknowledgments}

%
%

\begin{appendix}
\section{Related functions in factorization formulas}
\label{sec:app2}
	
The hard functions $h_i$ in the decay amplitudes come from the Fourier transformations
of quark and gluon propagators in the hard kernel. The explicit expressions are displayed
as follows,
\beq
h_{fe}(x_{1},x_{3},b_{1},b_{3})= && K_{0}\biggl (\sqrt{x_{1}x_{3}(1-r_{2}^{2})}m_{B^0}b_{1}\biggl)
\biggl[\theta (b_{1}-b_{3})K_{0}\biggl(\sqrt{x_{3}(1-r_{2}^{2})}m_{B^0}b_{1}\biggl) \notag \\
&& \cdot I_{0}\biggl(\sqrt{x_{3}(1-r_{2}^{2})}m_{B^0}b_{3}\biggl)+\theta (b_{3}-b_{1})K_{0}
\biggl(\sqrt{x_{3}(1-r_{2}^{2})}m_{B^0}b_{3}\biggl) \notag \\
&& \cdot I_{0}\biggl(\sqrt{x_{3}(1-r_{2}^{2})}m_{B^0}b_{1}\biggl)\biggl]S_{t}(x_{3}),
\eeq
\beq
h_{nfe}(x_{1},x_{2},x_{3},b_{1},b_{2}) = && \biggl\{\theta(b_{2}-b_{1}) I_{0}(m_{B^0}
\sqrt{x_{1}x_{3}(1-r_{2}^{2})}b_{1})K_{0}(m_{B^0}\sqrt{x_{1}x_{3}(1-r_{2}^{2})}b_{2})\hspace{0.5cm} \notag \\
&& + \ (b_{1}\leftrightarrow b_{2})\biggl\}\cdot \biggl(\begin{array}{cc} K_{0}(m_{B^0}F_{(1)}b_{2}),
	& {\rm for} \;\;  F_{(1)}^{2}>0 \\ \frac{\pi i}{2}H_{0}^{(1)}(m_{B^0}\sqrt{|F_{(1)}|^{2}}b_{2}),
	& {\rm for} \;\;  F_{(1)}^{2}<0
\end{array}
\biggl),
\eeq
where $J_{0}$ is the Bessel function, $K_{0}$ and $I_{0}$ are the modified Bessel functions.
The $F_{(1)}^{2}$ is defined by
\beq
F_{(1)}^{2}=(x_2-x_1)((x_3-x_2)r_2^2-x_3)+r_c^2.
\eeq
with $r_c = m_c/m_{B^0}$.

The expressions for the evolution functions $E_i(t)$ are defined as follows,
\beq
E_{fe}(t) &=& \alpha_s(t) \cdot \exp{[-S_{ab}(t)]}\cdot S_t(x) \;, \\
E_{nfe}(t) &=& \alpha_s(t) \cdot \exp{[-S_{cd}(t)]}\;,
\eeq
in which the jet function $S_{t}(x)$ arising from threshold resummation is universal
and has been parameterized in a simplified form
independent of decay channels, twist, and flavors as~\cite{Li:2001ay, Li:2002mi}
\beq
S_{t}(x)=\frac{2^{1+2c}\Gamma(3/2+c)}{\sqrt{2}\Gamma(1+c)}[x(1-x)]^{c},
\eeq
with $c = 0.4$~\cite{Li:2008tk}. This factor is normalized to unity.
And the Sudakov factors $S_{ab}(t)$ and $S_{cd}(t)$ used in this paper are
given as the following,
\beq
S_{ab}(t)&= & s(x_{1}P_{1}^{+},b_{1})+ s(x_{3}P_{3}^{-},b_{3})+s((1-x_{3})P_{3}^{-},b_{3})
-\frac{1}{\beta_{1}}\biggl[\ln \frac{\ln(t/\Lambda)}{-\ln(b_{1}\Lambda)}+\ln \frac{\ln(t/\Lambda)}
{-\ln(b_{3}\Lambda)}\biggl],
\eeq
\beq
S_{cd}(t)&=&s(x_{1}P_{1}^{+},b_{1})+ s_c(x_{2}P_{2}^{+},b_{2})+s_c((1-x_{2})P_{2}^{+},b_{2})
+ s(x_{3}P_{3}^{-},b_{1})+s((1-x_{3})P_{3}^{-},b_{1})
\non
&& \hspace{2cm} -\frac{1}{\beta_{1}}\biggl[2\ln \frac{\ln(t/\Lambda)}{-\ln(b_{1}\Lambda)}+\ln \frac{\ln(t/\Lambda)}{-\ln(m_c\Lambda)}\biggl].
\eeq
where the functions $s(q,b)$ and $s_c(q,b)$ could be found easily in Refs.~\cite{Liu:2020upy,Keum:2000wi,Lu:2000em}.
And the running hard scale $t_{i}^{,}s$ in the above equations are chosen as
the maximum energy scale to kill the large logarithmic radiative corrections
and they are given as follows,
\beq
t_{a} &=& \max(\sqrt{x_{3}(1-r_{2}^{2})}m_{B^0},1/b_{1},1/b_{3}),
\non
t_{b} &=& \max(\sqrt{x_{1}(1-r_{2}^{2})}m_{B^0},1/b_{1},1/b_{3}),  \\
t_{nfe} &=& \max(\sqrt{x_{1}x_{3}(1-r_{2}^{2})}m_{B^0},\sqrt{\left|(x_2-x_1)[(x_3-x_2)r_2^2-x_3]+r_c^2\right|}
\ m_{B^0},1/b_{1},1/b_{2}). \nonumber
\eeq

\end{appendix}

%
%

\end{document}